\documentclass[twocolumn]{aastex631}
\DeclareSymbolFont{matha}{OML}{txmi}{m}{it}
\DeclareMathSymbol{\varv}{\mathord}{matha}{118}

\usepackage{tabularx}
\usepackage{natbib}
\usepackage{amsmath}
\usepackage{amssymb}
\usepackage{bm}
\usepackage{comment}
\usepackage{hyperref}
\usepackage[normalem]{ulem}
\usepackage{soul}
\usepackage{tablefootnote}
\usepackage{graphicx}
\usepackage{color, colortbl}

\newcommand{\Athena}{{\sc Athena}}
\newcommand{\rhog}{\rho_{g}}
\newcommand{\rhop}{\rho_{p}}
\newcommand{\vecx}{\textit{\textbf{x}}}
\newcommand{\vecy}{\textit{\textbf{y}}}
\newcommand{\vecz}{\textit{\textbf{z}}}
\newcommand{\vecu}{\textit{\textbf{u}}}
\newcommand{\vecv}{{\bm{\varv}}}
\newcommand{\vecI}{\textit{\textbf{I}}}
\newcommand{\vecOmega}{\boldsymbol{\Omega}}

\newcommand{\Zcrit}{Z_{\rm{crit}}}
\newcommand{\Npar}{N_{\rm{par}}}

\newcommand{\JLsays}[1]{{\bf \color{blue}[Jay: #1]}}
\newcommand{\jakesays}[1]{{\bf \color{red}[JBS: #1]}}

\definecolor{brown}{rgb}{0.42,0.24,0.07}

\defcitealias{carrera_how_2015}{C15}
\defcitealias{Yang2017}{Y17}
\defcitealias{LiYoudin21}{LY21}

\usepackage{CJKutf8}
\newcommand{\ctext}[1]{\textup{\begin{CJK*}{UTF8}{bkai}#1\ignorespacesafterend\end{CJK*}}}

\newcommand\textkorean[1]{%
\begin{CJK}{UTF8}{mj}#1\ignorespacesafterend\end{CJK}}

\begin{document}

\title{Probing Conditions for Strong Clumping by the Streaming Instability: Small Dust Grains and Low Dust-to-gas Density Ratio}

\correspondingauthor{Jeonghoon Lim}
\email{jhlim@iastate.edu}

\author[0000-0003-2719-6640]{Jeonghoon Lim (\textkorean{임정훈})
}
\affiliation{Department of Physics and Astronomy, Iowa State University, Ames, IA 50010, USA}

\author[0000-0002-3771-8054]{Jacob B. Simon}
\affiliation{Department of Physics and Astronomy, Iowa State University, Ames, IA 50010, USA}

\author[0000-0001-9222-4367]{Rixin Li
\begin{CJK*}{UTF8}{gkai}(李日新)\end{CJK*}
}
\altaffiliation{51 Pegasi b Fellow}
\affiliation{Department of Astronomy, University of California, Berkeley, Berkeley, CA 94720, USA}

\author[0000-0001-6259-3575]{Daniel Carrera}
\affiliation{Department of Physics and Astronomy, Iowa State University, Ames, IA 50010, USA}
\affiliation{New Mexico State University, Department of Astronomy, PO Box 30001 MSC 4500, Las Cruces, NM 88001, USA}

\author[0000-0003-0412-760X]{Stanley A. Baronett}
\affiliation{Department of Physics and Astronomy, University of Nevada, Las Vegas, Box 454002, 4505 S. Maryland Pkwy., Las Vegas, NV 89154-4002, USA}

\author[0000-0002-3644-8726]{Andrew N. Youdin}
\affiliation{Department of Astronomy and Steward Observatory, University of Arizona, Tucson, Arizona 85721, USA}
\affiliation{The Lunar and Planetary Laboratory, University of Arizona, Tucson, Arizona 85721, USA}

\author[0000-0002-3768-7542]{Wladimir Lyra}
\affiliation{New Mexico State University, Department of Astronomy, PO Box 30001 MSC 4500, Las Cruces, NM 88001, USA}

\author[0000-0003-2589-5034]{Chao-Chin Yang (\ctext{楊朝欽})}
\affiliation{Department of Physics and Astronomy, The University of Alabama,
    Box~870324, Tuscaloosa, AL~35487-0324, USA}

\begin{abstract}
The streaming instability (SI) is a leading mechanism for concentrating solid particles into regions dense enough to form planetesimals. Its efficiency in clumping particles depends primarily on the dimensionless stopping time ($\tau_s$, a proxy for particle size) and dust-to-gas surface density ratio ($Z$). Previous simulations identified a critical $Z$ ($\Zcrit$) above which strong clumping occurs, where particle densities exceed the Hill density (thus satisfying a condition for gravitational collapse), over a wide range of $\tau_s$. These works found that for $\tau_s \leq 0.01$, $\Zcrit$ was above the ISM value $(\sim 0.01)$. In this work, we reexamine the clumping threshold using 2D axisymmetric, stratified simulations at high resolution and with relatively large (compared to many previous simulations) domain sizes. Our main results are as follows: First, when $\tau_s = 0.01$, strong clumping occurs even at $Z \lesssim 0.01$, lower than $\Zcrit$ found in all previous studies. Consequently, we revise a previously published fit to the $\Zcrit$ curve to account for this updated $\Zcrit$. Second, higher resolution results in a thicker dust layer, which may result from other instabilities manifesting, such as the vertical shearing streaming instability. Third, despite this thicker layer, higher resolution can lead to strong clumping even with a lower midplane dust-to-gas density ratios (which results from the thicker particle layer) so long as $Z \gtrsim \Zcrit$. Our results demonstrate the efficiency of the SI in clumping small particles at $Z \sim 0.01$, which is a significant refinement of the conditions for planetesimal formation by the SI.

\end{abstract}

\keywords{Planet formation (1241), Protoplanetary disks (1300), Planetesimals (1259), Gas-to-dust ratio (638), Hydrodynamics (1963)}

\section{Introduction}\label{sec:intro}
Planet formation begins with the growth of micron-sized dust grains in a protoplanetary disk (PPD) and encompasses various stages of growth, typically distinguished by the size of the solid objects involved. Among these, the formation of kilometer-scale planetesimals, while an integral step in the birth of fully grown planets, remains one of the least well understood stages (we refer the reader to \citealt{Simon2022review} for a recent review). 

In particular, while collisional coagulation of micrometer dust grains efficiently produces millimeter- to centimeter-sized pebbles, growth of these pebbles to larger sizes is halted by several barriers. These barriers include the bouncing barrier \citep{Zsom10,DominikDullemond24}, the fragmentation barrier \citep{DominikTielens97,DullemondDominik05}, and the radial drift barrier \citep{Adachi76,Weidenschilling77}. These barriers pose a significant challenge to the formation of planetesimals. While there are works that suggest collisional growth can be aided in a number of ways (e.g., the growth of porous dust aggregates into larger bodies; \citealt{Okuzumi2012,Kataoka2013}; see \citealt{Simon2022review} for a more detailed discussion), another alternative to forming planetesimals is some ``top down" mechanism by which pebbles are concentrated into sufficiently high densities such that gravitational collapse occurs.

The streaming instability (SI; \citealt{YG05,johansen_rapid_2007}) is one of the leading mechanisms to produce planetesimals via this process. The SI arises from the angular momentum exchange between solid particles and the gas via aerodynamic coupling and results in spontaneous concentration of the solids into radially narrow filamentary regions of high particle densities \citep{BaiStone10b_stratified,YangJohansen14,carrera_how_2015,Yang2017,LiYoudin21}. High resolution simulations with particle self-gravity implemented have confirmed that planetesimals form within these dense filaments via gravitational collapse \citep{Johansen15,simon_mass_2016,Simon17,Schafer17,abod_mass_2019,Li_demographics_2019}. 

The SI has passed some tests imposed by observational constraints\footnote{The size distribution of planetesimals from SI simulations remains an open problem since it does not match the observed distribution \citep{Kavelaars2021} in terms of the total number of planetesimals and the shape of the distribution (see Section 4 of \citealt{Simon2022review} for a discussion of this problem).} (e.g., the properties of Kuiper Belt binaries; \citealt{Nesvorny2019,Nesvorny2021} and the densities of Kuiper Belt objects; \citealt{2024PSJ.....5...55C}). As such, it is important to investigate under what conditions the SI operates. More specifically, it is crucial to study how physical and numerical parameters affect the formation of dense particle filaments produced by the SI.  The two physical parameters that have been studied most extensively are the size of solid particles and their abundance relative to the gas. The former is typically parameterized by the dimensionless stopping time $(\tau_s)$, which determines the timescale (relative to the  dynamical time, which is defined by the inverse of the local Keplerian frequency) over which the relative velocity between particles and the gas exponentially decays due to the gas drag on the particles. The latter is quantified by the column density ratio of particles to the gas $(Z)$. Another physical parameter, though much less well-studied within the context of SI-induced clumping is the radial gas pressure gradient, which sets the length scales of the SI \citep{YG05,johansen_protoplanetary_2007,Baronett24}. 

Several previous works have investigated the the clumping boundary, i.e., determining the critical $Z$ values $(\Zcrit)$, above which the SI forms strong concentrations of solid particles, as a function of $\tau_s$ (\citealt[hereafter \citetalias{carrera_how_2015}]{carrera_how_2015}; \citealt[hereafter \citetalias{Yang2017}]{Yang2017}; \citealt[hereater \citetalias{LiYoudin21}]{LiYoudin21}). They found that strong particle concentration occurs when $Z \lesssim 0.01$ (the dust-to-gas ratio in the ISM) only when $\tau_s \sim 0.1$, but relatively high $Z$ ($Z \gtrsim 0.02$) is needed for strong concentration of particles if $\tau_s \leq 0.01$. 

A comparison between these previous works \citepalias{carrera_how_2015,Yang2017,LiYoudin21} suggests that $\Zcrit$ can be lowered by using higher grid resolutions and radially wider computational domains, the latter of which provides a larger particle mass reservoir and thus allows formation of multiple filaments and merging between these filaments (\citealt{YangJohansen14}; \citetalias{Yang2017}; \citealt{Li18}).
For example, \citetalias{Yang2017} found smaller $\Zcrit$ than \citetalias{carrera_how_2015} for $\tau_s = 0.001-0.01$  due to the aforementioned improvements as well as an improved numerical algorithm for handling the stiff drag force in certain regions of parameter space \citep{Yang_Johansen16}. More recently, \citetalias{LiYoudin21} significantly lowered $\Zcrit$ across a wide range of $\tau_s$ (i.e., $0.001 \leq \tau_s \leq 1$) compared to \citetalias{carrera_how_2015}. However, both \citetalias{Yang2017} and \citetalias{LiYoudin21} reported relatively similar $\Zcrit$ as was found previously ($\Zcrit \gtrsim 0.02)$ for $\tau_s \leq 0.01$. Given that \citetalias{Yang2017} used higher grid resolution but smaller domain sizes, while \citetalias{LiYoudin21} used larger domain sizes with a fixed grid resolution smaller than the highest resolution of \citetalias{Yang2017}, the threshold for $\tau_s \leq 0.01$ may be further reduced by combining both large domain sizes and high grid resolution.

Motivated by this, we reexamine the clumping boundary by performing 2D local axisymmetric simulations combining both high grid resolution and large computational domains. To make a more direct connection to \citetalias{LiYoudin21}, we focus on what is known as {\it strong clumping}, which is when the SI-induced concentrations have density that surpass the Hill density (see Section \ref{sec:methods:parameters} for details).   Our highest resolution is four times greater than the fiducial resolution used by \citetalias{LiYoudin21}, while maintaining the same radial extent as in their study. We find that $\Zcrit \lesssim 0.01$ (referring to the critical $Z$ above which {\it strong clumping} occurs) at $\tau_s=0.01$, which is lower than previously shown, whereas $\Zcrit$ at $\tau_s=0.001$ lies between 0.02 and 0.03, similar to that found previously \citepalias{Yang2017,LiYoudin21}.

Our paper is structured as follows. We describe our numerical methods and the initial conditions of our simulations in Section \ref{sec:method}. We present the results with a revised fit to the clumping threshold in Section \ref{sec:result:criticalZ}. We provide detailed analysis of our simulations with $\tau_s=0.01$ in Sections \ref{sec:result:RadialConcentration}-\ref{sec:result:HpEps}. In Section \ref{sec:discussion}, we compare our results to the linear theory of unstratified SI and to \citetalias{Yang2017} and \citetalias{LiYoudin21}. We provide the summary and conclusions in Section \ref{sec:summary}. 

\section{Method}\label{sec:method}

\subsection{Numerical Algorithm and Governing Equations}\label{sec:methods:algorithm}

We use the \Athena \ \citep{Stone08} code to run simulations of the SI by evolving both gas (which we assume to be isothermal and unmagnetized) and solid particles and accounting for their aerodynamically coupling to each other. We employ the local shearing box approximation \citep{Hawley95, stone_implementation_2010} in which our computational domain is treated as a small, co-rotating patch of the disk, centered on a fiducial radius $(R)$, and rotating with Keplerian angular frequency $(\Omega)$. Since the size of the shearing box is much smaller than $R$, the governing equations can be expanded into Cartesian coordinates $(x,y,z)$; the relationship between the local Cartesian and cylindrical ($r, \phi, z'$) coordinates is: $x=r-R$, $y=R\phi$, and $z = z'$. The shearing box is stratified such that the gas maintains hydrostatic equilibrium with vertical scale height $H$, whereas the particles settle toward the disk midplane. We apply the standard shearing periodic boundary conditions \citep{Hawley95} in $x$, and the modified outflow boundary conditions described in \cite{Li18} as our vertical boundary conditions (VBCs). In simulations with outflow VBCs, we renormalize the total mass of gas in the entire domain after each time step since the VBCs allow material to leave the domain \citep{Li18}. However, we check that the amount of mass being replenished per timestep in those simulations is negligible; the fractional decrease in the total mass before being renormalized is $\sim \mathcal{O}(10^{-7}$-$10^{-6})$ per timestep. Simulations in this work are 2D ($x$-$z$) axisymmetric.

The hydrodynamic fluid equations in the shearing box are numerically solved by the unsplit Corner Transport Upwind (CTU) integrator \citep{Colella1990, 2008JCoPh.227.4123G}, third-order spatial reconstruction \citep{Colella1984}, and an HLLC Riemann solver \citep{Toro06}. The Crank-Nicolson method is employed to integrate the shearing box source terms (see \citealt{stone_implementation_2010}). The hydrodynamic equations are written as follows: 

\begin{equation}\label{eq:gas:continuity}
    \frac{\partial{\rhog}}{\partial{t}} + \nabla \cdot (\rho_g \textit{\textbf{u}}) = 0,
\end{equation}

\begin{equation}\label{eq:gas:momentum}
    \begin{split}
    \frac{\partial{\rhog \vecu}}{\partial{t}} + \nabla \cdot (\rhog \vecu \vecu + P \vecI) & = \\
    \rhog \left[ 3\Omega^2 \vecx - \Omega^2 \vecz  +2 \vecu  \times \vecOmega \right] 
    & + \rhop \frac{\vecv - \vecu}{t_{\rm stop}},  
    \end{split}
\end{equation}

\begin{equation}\label{eq:gas:eos}
    P = \rhog c_s^2.
\end{equation}

\noindent
Here, $\rhog$ is the gas mass density, $\vecu$ is the gas velocity, $c_s$ is the isothermal sound speed, $\vecI$ is the identity matrix, and $P$ is the isothermal gas pressure defined in Equation (\ref{eq:gas:eos}). On the right hand side of the momentum equation (Equation \ref{eq:gas:momentum}), the terms in the square brackets represent radial tidal forces (gravitational and centrifugal), vertical gravity, and the Coriolis force, respectively. The last term is the back-reaction from the particles to the gas; $\rhop$, $\vecv$, and $t_{\rm{stop}}$ are the particle mass density, particle velocity, and stopping time of particles, respectively.  The isothermal equation of state (Equation~\ref{eq:gas:eos}) closes the system. We employ the orbital advection scheme \citep{masset_fargo_2000} in which the Keplerian shear $\vecu_K=-(3/2)\Omega x \hat{\vecy}$ is subtracted and integrated following \cite{stone_implementation_2010}, while $\vecu'=\vecu-\vecu_K$ is evolved numerically using the algorithms described above.  

The solid particles are treated as Lagrangian super-particles, each of which is a statistical representation of a much larger number of particles with the same physical properties. The following equation of motion for the particle $i$ (out of $N_{\textrm{par}}$ total particles) is solved with the semi-implicit integrator of \cite{BaiStone10a}.
\begin{equation}\label{eq:par:momentum}
    \begin{split}
    \frac{d \vecv_i}{dt} & =  3\Omega^2\vecx_i - \Omega^2\vecz_i + 2 \vecv_i \times \vecOmega - \frac{\vecv_i - \vecu}{t_{\rm stop}}  \\ 
    & - 2\eta \vecu_K \Omega \hat{\vecx}.
    \end{split}
\end{equation}
On the right-hand side of the equation, the first, second, and third terms are the radial tidal forces (again, gravitational and centrifugal), vertical gravity, and Coriolis force, respectively. The fourth term is the aerodynamic drag force on individual particles. The last term represents a constant inward acceleration of particles that causes inward radial drift, which results from a negative gas pressure gradient (\citealt{Johansen06}; see \cite{BaiStone10a} for \Athena ~implementation). The dimensionless parameter $\eta$ is the fraction of Keplerian velocity by which the orbital velocity of particles is effectively increased, whereas the gas stays at Keplerian in the absence of particle feedback. Even though this approach is different to what happens in real PPDs where the gas orbits with sub-Keplerian speed, while particles stay at Keplerian, it does not change the relevant physics (see \citealt{BaiStone10a} for details of the approach).

Similar to the gas, an orbital advection scheme is applied to subtract the Keplerian shear from the azimuthal velocity of the particles. This shear component is handled the same way as the gas (see above), and the departure from Keplerian $\vecv_i' = \vecv_i - \vecu_K$ is integrated via \cite{BaiStone10a}.

To calculate the drag force in Equation (\ref{eq:par:momentum}), we use the triangular-shaped cloud scheme (TSC, \citealt{BaiStone10a}) to interpolate the gas velocity to particle locations. The same method is used to map particle momenta from particle locations to grid cell centers in order to calculate the particle back-reaction in Equation (\ref{eq:gas:momentum}).

In our simulations with $\tau_s=0.001$ and 0.01, the coupled equations can easily become stiff in regions of high particle density (see e.g., \citealt{BaiStone10a,Yang_Johansen16} for the description of the stiffness). Following \citet{BaiStone10a}, the stiffness parameter in \Athena~is:
\begin{equation}\label{eq:stiffness}
    \chi \equiv \frac{(\rho_p/\rho_g) \Delta t_{\rm{CFL}}}{\rm{max}(\tau_s,\Delta t_{\rm{CFL}})},
\end{equation}
where $\Delta t_{\rm{CFL}} \sim C_0\Delta x/c_s$, $\Delta x$ is the width of a grid cell (see Section \ref{sec:method:size_res} for grid resolutions we use), and $C_0 = 0.4$ is the Courant-Friedrichs-Lewy (CFL) number in \Athena. When $\chi$ exceeds order unity, numerical instabilities may arise and result in unphysically large growth of particle and gas velocities \citep{BaiStone10a}. To avoid this, we adopt the same technique as developed by \citetalias{LiYoudin21} (see their Equation 23) that reduces the numerical time step to keep $\chi$ below a certain threshold value $(\chi_{\rm{th}})$. 

We found that most of the strong-clumping runs show unstable evolution in the absence of applying this technique. In these runs, we restart these ``stiff simulations" at a time when $\chi \sim \chi_{\rm{th}} = 1.0$ with the time step restriction applied. In the run with $\tau_s=0.001$, we use a stricter value of $\chi_{\rm{th}} = 0.75$ from the beginning of the simulation to be more conservative. Additionally, we switch to the fully-implicit particle integrator developed by \citet{BaiStone10a} in this particular simulation. In Table \ref{tab:tab1}, we label simulations where we applied the technique.

The length, time, and mass units are $H$, $\Omega^{-1}$, and $\rho_{g0}H^3$, respectively, where $\rho_{g0}$ is initial gas density at the midplane. In our simulations, $H=\Omega=\rho_{g0}=1$.

\subsection{Simulation Details} \label{sec:methods:parameters}
Table \ref{tab:tab1} lists simulations in this work with all relevant details. We explain those details in the following subsections. 

\subsubsection{Dimensionless Parameters}
In numerical simulations of the SI in the presence of vertical gravity (but without particle self-gravity), $\tau_s$, $Z$, and $\Pi$ are the three fundamental dimensionless parameters that influence the evolution of the SI. The first of these, $\tau_s$, parameterizes the degree of aerodynamic coupling between the gas and particles: 
\begin{equation} \label{eq:taus}
    \tau_s = t_{\rm{stop}}\Omega,
\end{equation}
where $t_{\rm{stop}}$ is proportional to the particle size (see e.g, equation (1.48a) in \citealt{2013pss3.book....1Y} for the formula for $t_{\rm{stop}}$). We consider four $\tau_s$ values: 0.001, 0.01, 0.02, and 0.1.

The abundance of solid particles relative to the gas is quantified by the ratio of surface densities of the particles $(\Sigma_p)$ to the gas $(\Sigma_g)$:
\begin{equation} \label{eq:surfratioZ}
    Z = \left \langle \frac{\Sigma_{p0}}{\Sigma_{g0}} \right\rangle.
\end{equation}
Here, $\Sigma_{p0} (\Sigma_{g0})$ is the initial surface density of particles (the gas), and $\langle \cdots \rangle$ denotes a spatial average. We select $Z$ values from the no-clumping runs (see below for the definition of no-clumping) of \citetalias{LiYoudin21} at a given $\tau_s$ and revisit whether or not strong clumping occurs but at higher resolutions than were previously used.

The global gas pressure gradient in the disk is parameterized by
\begin{equation} \label{eq:pressure_gradient}
    \Pi = \frac{\eta u_K}{c_s} = \frac{\eta r}{H}.
\end{equation}
This parameter quantifies the departure of the gas azimuthal velocity from Keplerian in units of $c_s$; furthermore, the SI length scale is on the order of $\eta r$ \citep{YG05}. In line with previous studies \citepalias{carrera_how_2015,Yang2017,LiYoudin21}, $\Pi = 0.05$ for all simulations. 

It is worth noting that \cite{sekiya_two_2018} showed that it is actually the ratio of these latter two quantities $Z/\Pi$ that determines the structure of SI-induced filaments; thus, it seems plausible that $Z/\Pi$ is a fundamental parameter and not $Z$ and $\Pi$ separately. However, for this paper, we change only $Z$, while fixing $\Pi$ at the value expressed above. We discuss this choice of parameter variation further in Section \ref{sec:result:criticalZ}.

As we neglect self-gravity in this work, we must set a criterion for planetesimal formation. We assume a disk with $Q=32$ (same as in \citetalias{LiYoudin21}), where $Q$ is the Toomre parameter \citep{1960AnAp...23..979S,Toomre1964}, to parameterize the Hill density
\begin{equation}\label{eq:hill}
    \rho_H= 9\sqrt{\frac{\pi}{8}}Q\rho_{g0} = 180\rho_{g0}.
\end{equation}
A simulation is regarded as a strong-clumping run if the maximum particle density surpasses $\rho_H$. By contrast, our no-clumping runs have time-averaged maximum particle densities that are 1--2 orders of magnitude smaller than $\rho_H$ (see Table \ref{tab:tab2} for the time-averages of these densities).

\subsubsection{Domain Size and Grid Resolution}\label{sec:method:size_res}
We set our fiducial radial domain size ($L_x$) to be $0.8H$ for simulations with $\tau_s \geq 0.01$ to allow for the formation of multiple filaments and to account for their interaction, which includes mergers of filaments and thus significant increases in the maximum particle density. Previous studies showed that when $\Pi=0.05$, $L_x\gtrsim 0.3H$ is needed to form more than one filament for $\tau_s \cong0.3$ \citep{YangJohansen14,Li18}. Thus, our choice of $L_x=0.8H$ is sufficient to have multiple filaments. For the simulation with $\tau_s=0.001$, we instead use $L_x = 0.2H$ due to this run's increased computational cost associated with stricter limits on the time step (to handle stiffness) compared with the $\tau_s=0.01$ simulation. However, we still find that three filaments form, with one merging with another, leaving two filaments at the end of the simulation. This is consistent with \citetalias{Yang2017} and \citetalias{LiYoudin21} who also found several filaments in their $\tau_s=0.001$ runs with $L_x=0.2H$.


The vertical domain size is $L_z = 0.2H$ for $\tau_s=0.01$ and $L_z = 0.4H$ for $\tau_s \neq 0.01$. These specific values of $L_z$ were chosen to account for the fact that the minimum particle scale height (as a function of $\tau_s$) occurs when $\tau_s=0.01$ (at least within the range $0.001 \leq \tau_s \leq 0.1$\footnote{According to \citetalias{LiYoudin21}, the particle scale height decreases with increasing $\tau_s$ when $0.001 \leq \tau_s \leq 0.01$, owing to the more rapid sedimentation at larger $\tau_s$. On the other hand, the scale height increases with $\tau_s$ from $\tau_s=0.01$ to 0.1, which is likely due to stronger particle stirring by these larger particles. See \citetalias{LiYoudin21} for more a detailed explanation of the dependence of the particle scale height on $\tau_s$.}; see \citetalias{LiYoudin21}), allowing us to use a smaller domain for this $\tau_s$ value. However, we ran an additional simulation with $L_z = 0.4H$ at $\tau_s=Z=0.01$ to check if using a different $L_z$ affects our results (see Section \ref{sec:discussion:comparison_to_previous_studies}). Although the taller box has a slightly smaller particle scale height (see Table \ref{tab:tab2}), strong clumping is found in both of the runs, and thus our threshold for strong clumping remains the same.  In the case of $\tau_s = 0.1$, we find that using $L_z = 0.2H$ results in a strongly asymmetric (about the midplane) particle layer (see Appendix \ref{sec:appendixB:IssueOutflow} for details of this issue), further justifying our use of a larger domain for $\tau_s = 0.1$.


We consider three different grid resolutions for $\tau_s=0.01$, which are $1280/H$ (i.e., 1280 grid cells per $H$, or $\Delta x = H/1280$), $2560/H$, and $5120/H$; the smallest resolution is the fiducial resolution of \citetalias{LiYoudin21}. For $\tau_s = 0.001$ ($\tau_s > 0.01$), we run simulations only at a resolution of $5120/H$ ($2560/H$). We explain these choices in detail in Section \ref{sec:result:criticalZ}.

\subsubsection{Particle Resolution}\label{sec:method:particle_res}

The total number of particles is defined by:
\begin{equation}
	\Npar = n_pN_{\textrm{cell}} = n_pN_xN_z,
\end{equation}
where $n_p$ is the average number of particles per grid cell. We use $n_p$ of 1.0 in simulations with $L_z=0.2H$, while $n_p=0.5$ in those with $L_z=0.4H$. This is to keep the same $\Npar$ in the two types of simulations (though $\Npar$ changes if resolution and/or $L_x$ changes). We note that in stratified simulations, the settling of particles leads to particle-free cells above and below the particle layer that do not contribute to the particle resolution.  Therefore, we opt to use the effective particle resolution $(n_p^*$), which is more relevant near the midplane and is defined by:
\begin{equation}\label{eq:np*}
    n_p^*=n_p\left(\frac{L_z}{6H_p}\right). 
\end{equation}
We measure $n_p^*$ within $\pm 3H_p$ since most of particles are expected to be within the thickness. We report time-averaged values of $n_p^*$ during the pre-clumping phase (see Section \ref{sec:result} for the definition of the pre-clumping phase) in Table \ref{tab:tab2}, showing that $n_p^* > n_p$ in every simulation.

\subsubsection{Initialization}
The gas is initialized in vertical hydrostatic balance as a Gaussian density profile with scale height $H$. Similarly, particles are initially distributed with a Gaussian density profile in the vertical direction, and the initial vertical scale height $(H_{p0})$ is $\eta r/2$.  The initial midplane densities of gas and particles ($\rho_{p0})$ are given by: 
\begin{equation}\label{eq:rhog0}
    \rho_{g0} = \frac{\Sigma_{g0}}{\sqrt{2\pi} H},
\end{equation}
and 
\begin{equation}\label{eq:rhop0}
    \rho_{p0} = \frac{\Sigma_{p0}}{\sqrt{2\pi} H_{p0}},
\end{equation}

\noindent
respectively. Thus, $\langle \rho_{p0}/\rho_{g0} \rangle=2Z/\Pi$. Since we confirm that the vertical profile of the particle density is nearly Gaussian (at all times) in all simulations considered, we measure the particle scale height ($H_p$) at a given time by calculating the standard deviation of vertical positions ($z$) of the particles:

\begin{equation}\label{eq:hp}
    H_p = \sqrt{\frac{1}{(\Npar-1)}\sum_{i=1}^{\Npar}(z_i-\overline{z_i})^2}.
\end{equation}

The vertical velocities of the gas and the particles are initially zero. In the radial direction, the particles' positions are randomly chosen from a uniform distribution. We apply the Nakagawa—Sekiya—Hayashi (NSH) equilibrium solutions \citep{Nakagawa1986} to the horizontal velocities (i.e., radial and azimuthal) of the gas and the particles.  

We evolve the no-clumping runs of $\tau_s=0.001$ and 0.01 for at least $\sim 10^{4}\Omega^{-1}$. However some strong-clumping runs experience numerical instability associated with high stiffness; this manifests as spikes in the temporal evolution of the maximum particle density, similar to Figure 9 of \citetalias{LiYoudin21}. One way to mitigate this stiffness is to reduce the time step, as we described above. However, with this time step reduction, these particular runs become very computationally expensive, and it makes more sense to halt their integration once strong clumping has been achieved. We evolve simulations with $\tau_s > 0.01$ for shorter lengths of time: up to $6000\Omega^{-1}$ and $3000\Omega^{-1}$ for $\tau_s=0.02$ and 0.1, respectively; we do this because larger $\tau_s$ generally produces strong clumping more rapidly (i.e., larger $\tau_s$ settle faster, which establishes an equilibrium layer faster, thus allowing more rapid strong clumping;  see also \citetalias{Yang2017}).

\begin{figure*}
    \centering
    \includegraphics[width=\textwidth]{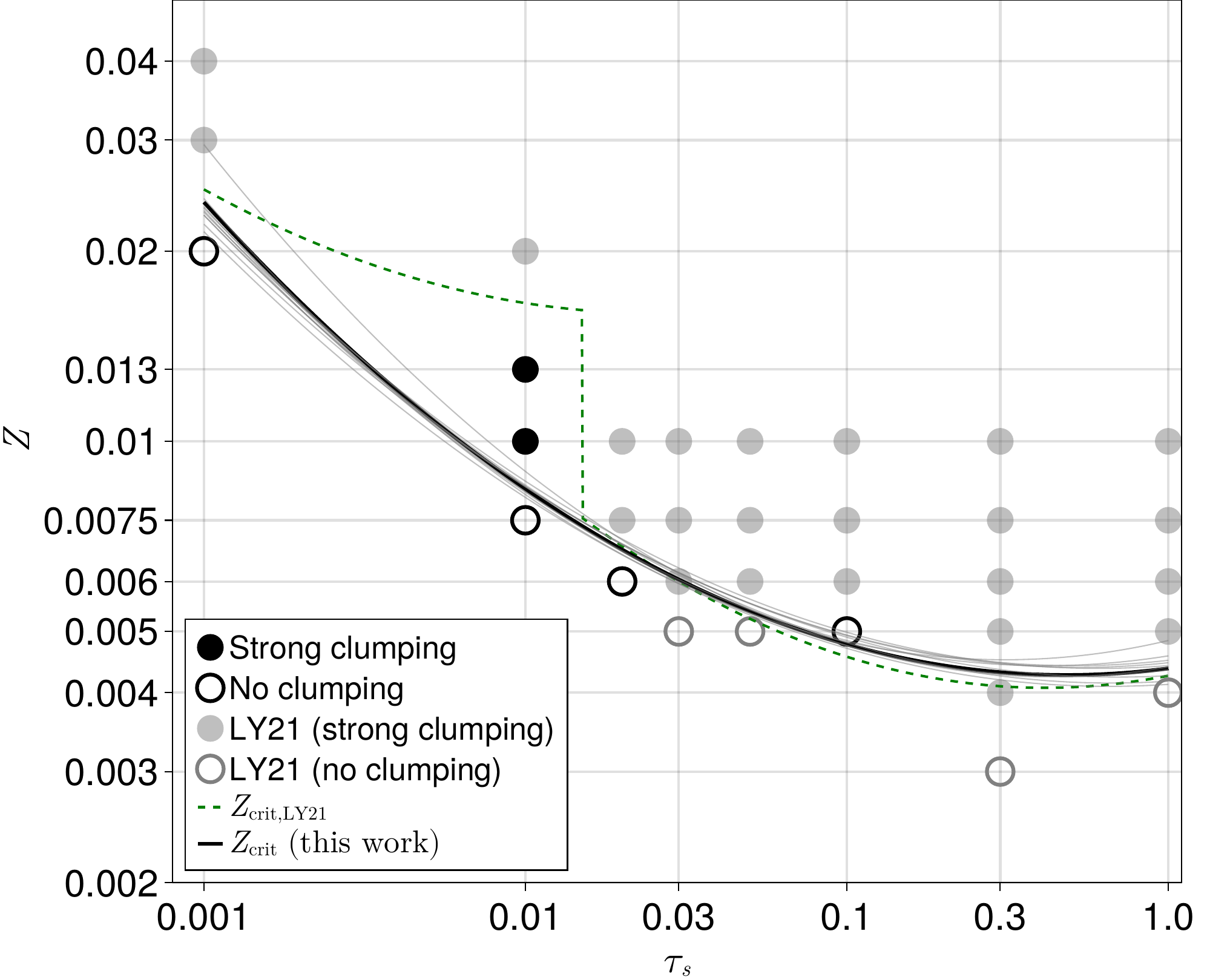}
    \caption{Summary of our simulations in terms of whether each simulation shows strong clumping (filled circles) or no clumping (empty circles). Simulations in this work are marked as black circles, whereas those in \citetalias{LiYoudin21} are marked as gray circles. The black curve is the best fit by least squares showing the minimum $Z$ value required for strong clumping (i.e., $\Zcrit$) as a function of $\tau_s$. The thin gray curves are random fits drawn from a normal distribution of the fitted parameters $A, ~B, ~C$ (Equation \ref{eq:Zcrit}). The $\Zcrit$ curve reported by \citetalias{LiYoudin21} is shown as the green dashed curve. Simulations in this work find {\it strong clumping at lower $Z$ values} than those in \citetalias{LiYoudin21} at $\tau_s=0.01$.}
    \label{fig:fig-Zcrit}
\end{figure*}

\begin{deluxetable*}{cccccccc}
\tablecaption{List of Simulations}\label{tab:tab1}
\tablehead{
\colhead{$\tau_s$} & \colhead{$Z$} & \colhead{$L_x\times L_z$} &  \colhead{Resolution} & \colhead{VBCs} & \colhead{$N_{\rm{par}}$} & \colhead{$t_{\rm{end}}$} & \colhead{Strong Clumping?} \\
\colhead{} &
\colhead{} &
\colhead{($H^2$)} &
\colhead{($H^{-1}$)} &
\colhead{}  &
\colhead{}  &
\colhead{($\Omega^{-1}$)} &
\colhead{} \\
\colhead{(1)} &
\colhead{(2)} &
\colhead{(3)} &
\colhead{(4)} &
\colhead{(5)} &
\colhead{(6)} &
\colhead{(7)} &
\colhead{(8)} 
}
\startdata
0.001\tablenotemark{a} & 0.02 & $0.2\times 0.4$\tablenotemark{b} & 5120 & outflow & $\sim 1.0\times 10^{6}$ & 10000 & N \\
0.01 & 0.0075 & $0.8\times 0.2$ & 1280 & outflow & $\sim 2.6\times 10^{5}$ & 14000 & N\\
0.01 & 0.0075 & $0.8\times 0.2$ & 2560 & outflow & $\sim 1.0\times 10^{6}$ & 14000 & N\\
0.01 & 0.0075 & $0.8\times 0.2$ & 5120 & outflow & $\sim 4.2\times 10^{6}$ & 10079 & N\\
0.01 & 0.01 & $0.8\times 0.2$ & 1280 & outflow & $\sim 2.6\times 10^{5}$   & 14000 & N\\
0.01 & 0.01 & $0.2\times 0.2$ & 2560 & outflow & $\sim 2.6\times 10^{5}$   & 14000 & N\\
\textbf{0.01\tablenotemark{a}} & \textbf{0.01} & \boldsymbol{$0.4\times 0.2$} & \textbf{2560} & \textbf{outflow} & \boldsymbol{$\sim 5.2\times 10^{5}$}   & \textbf{9761} & \textbf{Y}\\
\textbf{0.01} & \textbf{0.01} & \boldsymbol{$0.8\times 0.2$} & \textbf{2560} & \textbf{outflow} & \boldsymbol{$\sim 1.0\times 10^{6}$}   & \textbf{9168} & \textbf{Y}\\
0.01 & 0.01 & $0.8\times 0.2$ & 2560 & periodic & $\sim 1.0\times 10^{6}$  & 14000 & N\\
\textbf{0.01\tablenotemark{a}} & \textbf{0.01} & \boldsymbol{$0.8\times 0.4$} & \textbf{2560} & \textbf{outflow} & \boldsymbol{$\sim 1.0\times 10^{6}$}   & \textbf{11358} & \textbf{Y}\\
\textbf{0.01\tablenotemark{a}} & \textbf{0.01} & \boldsymbol{$0.4\times 0.4$} & \textbf{2560} & \textbf{periodic} & \boldsymbol{$\sim 5.2\times 10^{5}$} & \textbf{3278} & \textbf{Y}\\
\textbf{0.01\tablenotemark{a}} & \textbf{0.01} & \boldsymbol{$0.8\times 0.4$} & \textbf{2560} & \textbf{periodic} & \boldsymbol{$\sim 1.0\times 10^{6}$}  & \textbf{2740} & \textbf{Y}\\
\textbf{0.01\tablenotemark{a}} & \textbf{0.01} & \textbf{$0.8\times 0.2$} & \textbf{5120} & \textbf{outflow} & \boldsymbol{$\sim 4.2\times 10^{6}$}   & \textbf{6113} & \textbf{Y}\\
0.01 & 0.013 & $0.8\times 0.2$ & 1280 & outflow & $\sim 2.6\times 10^{5}$  & 14000 & N\\
\textbf{0.01} & \textbf{0.013} & \boldsymbol{$0.8\times 0.2$} & \textbf{2560} & \textbf{outflow} & \boldsymbol{$\sim 1.0\times 10^{6}$}  & \textbf{13736} & \textbf{Y}\\
\textbf{0.01\tablenotemark{a}} & \textbf{0.013} & \boldsymbol{$0.8\times 0.2$} & \textbf{5120} & \textbf{outflow} & \boldsymbol{$\sim 4.2\times 10^{6}$}  & \textbf{7158} & \textbf{Y}\\
0.02 & 0.006 & $0.8\times 0.4$ & 2560 & outflow & $\sim 1.0\times 10^{6}$  & 6000 & N\\
0.1 & 0.005 & $0.8\times 0.4$ & 2560 & outflow & $\sim 1.0\times 10^{6}$   & 3000 & N\\
0.1 & 0.005 & $0.8\times 0.2$ & 2560 & periodic & $\sim 1.0\times 10^{6}$  & 3000 & N\\
0.1 & 0.005 & $0.8\times 0.4$ & 2560 & periodic & $\sim 1.0\times 10^{6}$  & 3000 & N\\
\enddata
\tablecomments{ Columns: (1) dimensionless stopping time of particles (see Equation \ref{eq:taus}); (2) surface density ratio of particles to gas (see Equation \ref{eq:surfratioZ}); (3) domain size in units of gas scale height; (4) number of grid cells per gas scale height; (5) vertical boundary condition; (6) total number of particles; (7) total simulation time; (8) Whether or not strong clumping occurs (Y for yes and N for no). All runs have a radial pressure gradient of $\Pi = 0.05$. Strong-clumping runs are highlighted by bold-face.
}    
\tablenotetext{a}{In these simulations, we adopt the technique developed by \citetalias{LiYoudin21} (see their  Equation 23) to mitigate high stiffness by reducing the time step. For strong-clumping runs, we terminate simulations after strong clumping begins due to the prohibitive computational cost. We terminate the run with $\tau_s=0.001$ at $t\Omega=10^{4}$ even though no strong clumping occurred. This is because maximum particle density saturates around $10\rho_{g0}$ after $t\Omega \sim 8000$, and the computational cost is too high to run it further.}
\tablenotetext{b}{We use a smaller box size in this run to mitigate the added expense of a reduced time step due to high stiffness.}
\end{deluxetable*}

\begin{deluxetable*}{cccccccccccc}
\tablecaption{Time-averaged quantities}\label{tab:tab2}
\tablehead{
\colhead{$\tau_s$} & \colhead{$Z$} & \colhead{$L_x\times L_z$} &  \colhead{Resolution} & \colhead{VBCs} & \colhead{$\overline{\rho_{p,\rm{max}}}$} & \colhead{$\overline{H_p}$} & \colhead{$\overline{\epsilon}$} & \colhead{$\overline{n_p^*}$} & \colhead{$\overline{\rho_{p,95}}$} & \colhead{$\overline{\rho_{p,99}}$}  & \colhead{[$t_s,~t_e$]} \\
\colhead{} &
\colhead{} &
\colhead{$H^2$} &
\colhead{$H^{-1}$} &
\colhead{}  &
\colhead{$\rho_{g0}$} &
\colhead{$H$} & 
\colhead{} &
\colhead{} & 
\colhead{$\rho_{g0}$} &
\colhead{$\rho_{g0}$} &
\colhead{$\Omega^{-1}$} \\
\colhead{(1)} &
\colhead{(2)} &
\colhead{(3)} &
\colhead{(4)} &
\colhead{(5)} &
\colhead{(6)} &
\colhead{(7)} &
\colhead{(8)} &
\colhead{(9)} &
\colhead{(10)} &
\colhead{(11)} & 
\colhead{(12)} 
}
\startdata
0.001 & 0.02 & $0.2\times 0.4$ & 5120 & outflow & 8.5376 & 0.0071 & 2.8173 & 4.6948 & 3.6431  & 5.1707 & [4000,10000] \\
0.01 & 0.0075 & $0.8\times 0.2$ & 1280 & outflow & 5.0112 & 0.0078 &  0.9679 & 4.2735 & 1.3924 &  2.4548  &  [300, 14000]\\
0.01 & 0.0075 & $0.8\times 0.2$ & 2560 & outflow & 2.2174  & 0.0109 &  0.6869 & 3.0581 & 0.8576 & 1.1128  & [300, 14000]\\
0.01 & 0.0075 & $0.8\times 0.2$ & 5120 & outflow & 1.5125 & 0.0125 &  0.5977 & 2.6667 & 0.7145  & 0.8804  & [300, 10079]\\
0.01 & 0.01 & $0.8\times 0.2$ & 1280 & outflow & 6.2973 &  0.0068 &   1.4746 & 4.9022 & 2.1655  & 3.3562  & [300, 14000]\\
0.01 & 0.01 & $0.2\times 0.2$ & 2560 & outflow & 13.9340  & 0.0090 &   1.1069 & 3.7037 & 1.2608 &  4.7910 & [300, 14000] \\
\textbf{0.01} & \textbf{0.01} & \boldsymbol{$0.4\times 0.2$} & \textbf{2560} & \textbf{outflow} & \textbf{14.7816} & \textbf{0.0090} &  \textbf{1.1138} & \textbf{3.7037} & \textbf{1.4114}  &  \textbf{4.1952}   & \textbf{[300, 9333]}\\
\textbf{0.01} & \textbf{0.01} & \boldsymbol{$0.8\times 0.2$} & \textbf{2560} & \textbf{outflow} & \textbf{17.1261} & \textbf{0.0091} &  \textbf{1.0952} & \textbf{3.6630} & \textbf{1.3406} & \textbf{4.4084} & \textbf{[300, 8473]}\\
0.01 & 0.01 & $0.8\times 0.2$ & 2560 & periodic & 14.9996 &  0.0142 &  0.7029 & 2.3474 & 0.9462 & 2.6900  & [3000, 14000]\\
\textbf{0.01} & \textbf{0.01} & \boldsymbol{$0.8\times 0.4$} & \textbf{2560} & \textbf{outflow} & \textbf{10.6294}  & \textbf{0.0077}  &  \textbf{1.3011} & \textbf{4.3290} & \textbf{1.9022}  & \textbf{3.9331} &  \textbf{[300, 11257]}\\
\textbf{0.01\tablenotemark{a}} & \textbf{0.01} & \boldsymbol{$0.4\times 0.4$} & \textbf{2560} & \textbf{periodic} & \boldsymbol{$\cdots$} & \boldsymbol{$\cdots$} & \boldsymbol{$\cdots$} & \boldsymbol{$\cdots$} & \boldsymbol{$\cdots$} & \boldsymbol{$\cdots$} & \boldsymbol{$\cdots$} \\
\textbf{0.01\tablenotemark{a}} & \textbf{0.01} & \boldsymbol{$0.8\times 0.4$} & \textbf{2560} & \textbf{periodic} & \boldsymbol{$\cdots$} & \boldsymbol{$\cdots$} & \boldsymbol{$\cdots$} & \boldsymbol{$\cdots$} & \boldsymbol{$\cdots$} & \boldsymbol{$\cdots$} & \boldsymbol{$\cdots$}\\
\textbf{0.01} & \textbf{0.01} & \boldsymbol{$0.8\times 0.2$} & \textbf{5120} & \textbf{outflow} & \textbf{16.3287} & \textbf{0.0112} & \textbf{0.8949} & \textbf{2.9672} & \textbf{1.0444} &  \textbf{2.3817} & \textbf{[300, 5956]}\\
0.01 & 0.013 & $0.8\times 0.2$ & 1280 & outflow & 8.5238 & 0.0064 & 2.0402 & 5.2083 & 2.9962 & 4.4752 & [300, 14000]\\
\textbf{0.01} & \textbf{0.013} & \boldsymbol{$0.8\times 0.2$} & \textbf{2560} & \textbf{outflow} & \textbf{15.4770} &  \textbf{0.0085} & \textbf{1.5359} & \textbf{3.9216} & \textbf{2.1634} & \textbf{5.3415} & \textbf{[300, 12335]}\\
\textbf{0.01} & \textbf{0.013} & \boldsymbol{$0.8\times 0.2$} & \textbf{5120} & \textbf{outflow} & \textbf{26.2934} & \textbf{0.0102} & \textbf{1.2774} & \textbf{3.2680} & \textbf{1.5525}  & \textbf{5.6487} & \textbf{[300, 6944]}\\
0.02 & 0.006 & $0.8\times 0.4$ & 2560 & outflow & 1.6648 & 0.0128 &  0.4686 & 2.6042 &  0.6420 &  0.8278 & [200, 6000]\\
0.1 & 0.005 & $0.8\times 0.4$ & 2560 & outflow & 14.1626 &  0.0136  & 0.3674 & 2.4510 &   1.3551 &  2.6988 & [100, 3000]\\
0.1 & 0.005 & $0.8\times 0.4$ & 2560 & periodic & 12.8904  & 0.0197 &  0.2556 & 1.6920 & 1.1187  & 2.2739 & [100, 3000]\\
\enddata
\tablecomments{ Columns: (1) dimensionless stopping time of particles (see Equation \ref{eq:taus}); (2) surface density ratio of particles to gas (see Equation \ref{eq:surfratioZ}); (3) domain size in units of gas scale height; (4) the number of grid cells per gas scale height; (5) vertical boundary condition; (6) maximum particle density in units of the initial midplane gas density; (7) particle scale height (see Equation \ref{eq:hp}) in units of the gas scale height; (8) density ratio of particle-to-gas at the midplane (see Equation \ref{eq:eps}); (9) effective particle resolution (see Equation \ref{eq:np*}); (10)-(11) particle density at the 95th and 99th percentiles, respectively, in units of the initial midplane gas density;  (12) time interval over which quantities in columns (6)-(11) are averaged in units of $\Omega^{-1}$. Strong-clumping runs are highlighted by bold-face.}

\tablenotetext{a}{As the particle scale height never saturates before strong clumping happens, we opt not to provide time-averages for these runs.}
\end{deluxetable*}

\section{Results}\label{sec:result}
In this section, we present our results, mostly focusing on runs with $\tau_s=0.01$ where we find a {\it lower} threshold for strong particle clumping compared to previous works (\citetalias{carrera_how_2015,Yang2017,LiYoudin21}). We present our new critical $Z$ for strong clumping in Section \ref{sec:result:criticalZ}. In Section \ref{sec:result:RadialConcentration}, we address the radial concentration of particles for different $Z$ and grid resolutions. In Section \ref{sec:result:CharRhop}, we present histograms of the particle density to gauge the characteristic density of the SI-induced filaments. In Section \ref{sec:result:HpEps}, we investigate how $H_p$ and the midplane particle to gas density ratio change with grid resolution. Table \ref{tab:tab2} provides time-averaged quantities (denoted by $\overline{\cdots}$) of simulations considered in this work. We calculate time averages after the particles have settled from their initial positions but before the maximum density first exceeds (2/3)$\rho_H$ (termed as the pre-clumping phase, \citetalias{LiYoudin21}). In the case where strong clumping does not happen, the averaging is done up until the end of the simulations. All simulations of $\tau_s=0.01$ considered in this section have $(L_x,L_z)=(0.8,0.2)H$ and outflow VBCs. We discuss the effect of box size and boundary conditions in Section \ref{sec:discussion}.

\subsection{Critical $Z$ for Strong Particle Clumping}\label{sec:result:criticalZ}
Our primary result is shown in Figure~\ref{fig:fig-Zcrit}; we show a refined boundary for strong clumping based on our simulations, while also including a comparison with the most recent examination of this boundary by \citetalias{LiYoudin21}. Strong-clumping and no-clumping runs from our simulation suite are represented as filled and empty black circles, respectively. For the other $(\tau_s,Z)$ combinations that we did not explore, we adopt the results from \citetalias{LiYoudin21} and mark them as gray circles, again denoting strong (no) clumping as filled (empty) circles.  Table \ref{tab:tab1} shows the clumping results of our simulations.

We find strong clumping for $\tau_s=0.01$ at $Z=0.01$ and 0.013, which lowers the critical $Z$ $(\Zcrit)$ above which strong clumping occurs to below 0.01, which results in a smooth transition of the $\Zcrit$ curve between $\tau_s=0.01$ and 0.02. We present a detailed comparison between our work and \citetalias{LiYoudin21} in Section \ref{sec:discussion:comparison_to_previous_studies}. 



For simulations with $\tau_s \neq 0.01$, we motivate our specific grid resolutions based on previous studies that found that simulations with smaller $\tau_s$ require higher grid resolution for the SI to concentrate particles (\citealt{YangJohansen14}; \citetalias{Yang2017}). As such, we re-run some of the no-clumping runs of \citetalias{LiYoudin21} for $\tau_s \neq 0.01$ at higher resolution. For simulations with $\tau_s > 0.01$, we used a resolution of $2560/H$, assuming that the critical resolution, above which strong clumping occurs, for $\tau_s > 0.01$ is lower than that for $\tau_s = 0.01$ (i.e., $2560/H$). Since all of these remain no-clumping cases, we assume that resolutions $\gtrsim 1280/H$ do not alter the clumping results for $\tau_s > 0.01$.


We run the $\tau_s = 0.001$ case with the same box size as \citetalias{LiYoudin21} but with a higher resolution of $5120/H$, quadrupling their fiducial resolution, and for a longer run time of $10^{4}\Omega^{-1}$. While we cannot verify that running the simulation out further would still result in no clumping, we choose to end this simulation at this time due to high computational expense (running it to $10^{4}\Omega^{-1}$ cost $\sim 0.5$ million CPU hours) and classify it as no clumping. Apart from this, we emphasize that the no-clumping result may be due to the smaller box used and/or a critical resolution that is higher than ours. However, testing whether strong clumping would occur with larger box sizes or higher resolutions is beyond the scope of this investigation given the extreme expense of such simulations.\footnote{We estimate that using $L_x=0.8H$ at $5120/H$ would require at least 2 million CPU hours, and doubling the grid resolution (while keeping $L_x=0.2H$) would cost at least 4 million CPU hours.}


Since our runs are at higher resolution than \citetalias{LiYoudin21}, and since higher resolution is unlikely to quench strong clumping that has already manifested at lower resolution, we assume that the strong-clumping runs of \citetalias{LiYoudin21} for $\tau_s \neq 0.01$ will remain strong clumping at our resolution. This allows us to produce a $\Zcrit$ curve, as described below. That said, this assumption should be tested in future studies by conducting simulations of strong-clumping cases just above the $\Zcrit$ curve at similar resolutions to ours. 

In Figure~\ref{fig:fig-Zcrit}, the black and green curves show $\Zcrit$ as a function of $\tau_s$ for our results and those of \citetalias{LiYoudin21}, respectively. Both curves are quadratic in logarithmic space, and fit empirically by:

\begin{equation}\label{eq:Zcrit}
    \log{(\Zcrit)}=A(\log\tau_s)^2+B\log \tau_s+C.
\end{equation}
To obtain the coefficients $A, B, C$, we perform a least squares fit assuming that at a given $\tau_s$, the critical $Z$ value lies between adjacent empty (no clumping) and filled (strong clumping) circles. We confirm that using higher grid resolution does not change no-clumping cases to strong-clumping cases for $\tau_s \neq 0.01$; as such, we simply adopt the strong clumping results found by \citetalias{LiYoudin21} in the region of $(\tau_s,Z)$ parameter space that we have not explored in our work (i.e., gray circles in Figure \ref{fig:fig-Zcrit}) for the least squares fit. According to the the best fit, $A=0.10,~B=0.07,~C=-2.36$ (black curve), close to the \citetalias{LiYoudin21} curve (green) for $\tau_s > 0.015$ and $\tau_s = 0.001$. The thin gray curves are randomly drawn from a normal distribution of $(A,B,C)$ and thus represent the uncertainties in the best-fit curve.

We do not normalize our $\Zcrit$ by $\Pi$. While $Z/\Pi$ may be a fundamental parameter for stratified SI simulations (as opposed to $Z$ and $\Pi$ separately; see \citealt{sekiya_two_2018}), we caution that few simulations have been conducted to rigorously validate the result. Thus, further extensive simulations, varying $\tau_s$, $Z$, and $\Pi$ while maintaining a constant $Z/\Pi$, are required to confirm this approach.



\begin{figure*}
    \centering
    \includegraphics[width=\textwidth]{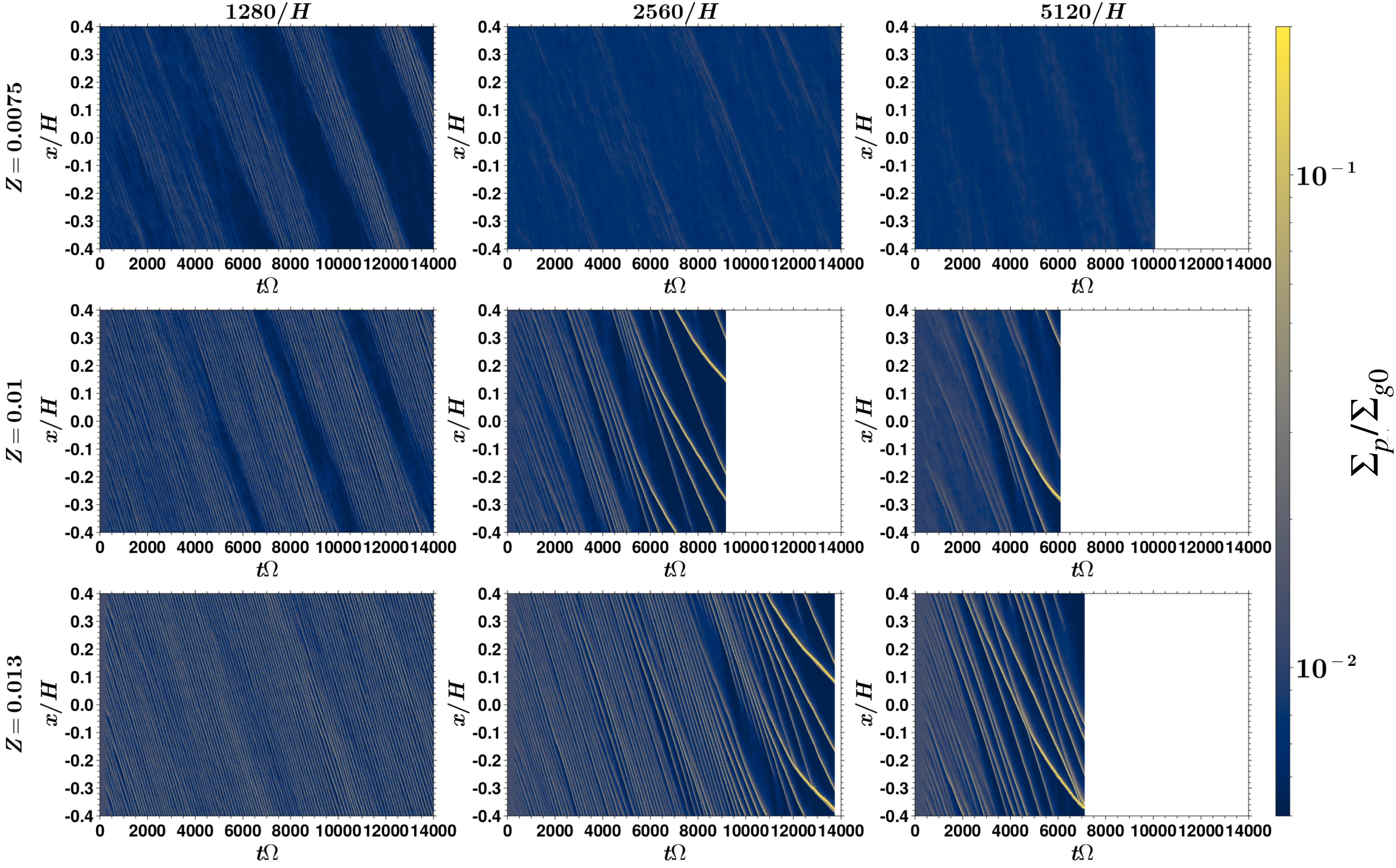}
    \caption{Space-time plots of the particle density integrated over $z$ ($\Sigma_p$). We plot the quantity vs. $x$ and time for runs with $\tau_s=0.01$. From top to bottom, $Z=0.0075$, 0.01, and 0.013, respectively. Grid resolution increases from $1280/H$ (left) to $5120/H$ (right). No dense filaments are seen in the $Z=0.0075$ cases, whereas filaments formed in the other two $Z$ values at resolutions higher than $1280/H$ are disrupted; this leads to either the feeding of downstream filaments $(Z=0.01)$ or filament mergers $(Z=0.013)$. Based on our simulations, both mechanisms facilitate strong clumping (see the text for more details).
    }
    \label{fig:fig-rhopt}
\end{figure*}

\begin{figure*}
    \includegraphics[width=\textwidth]{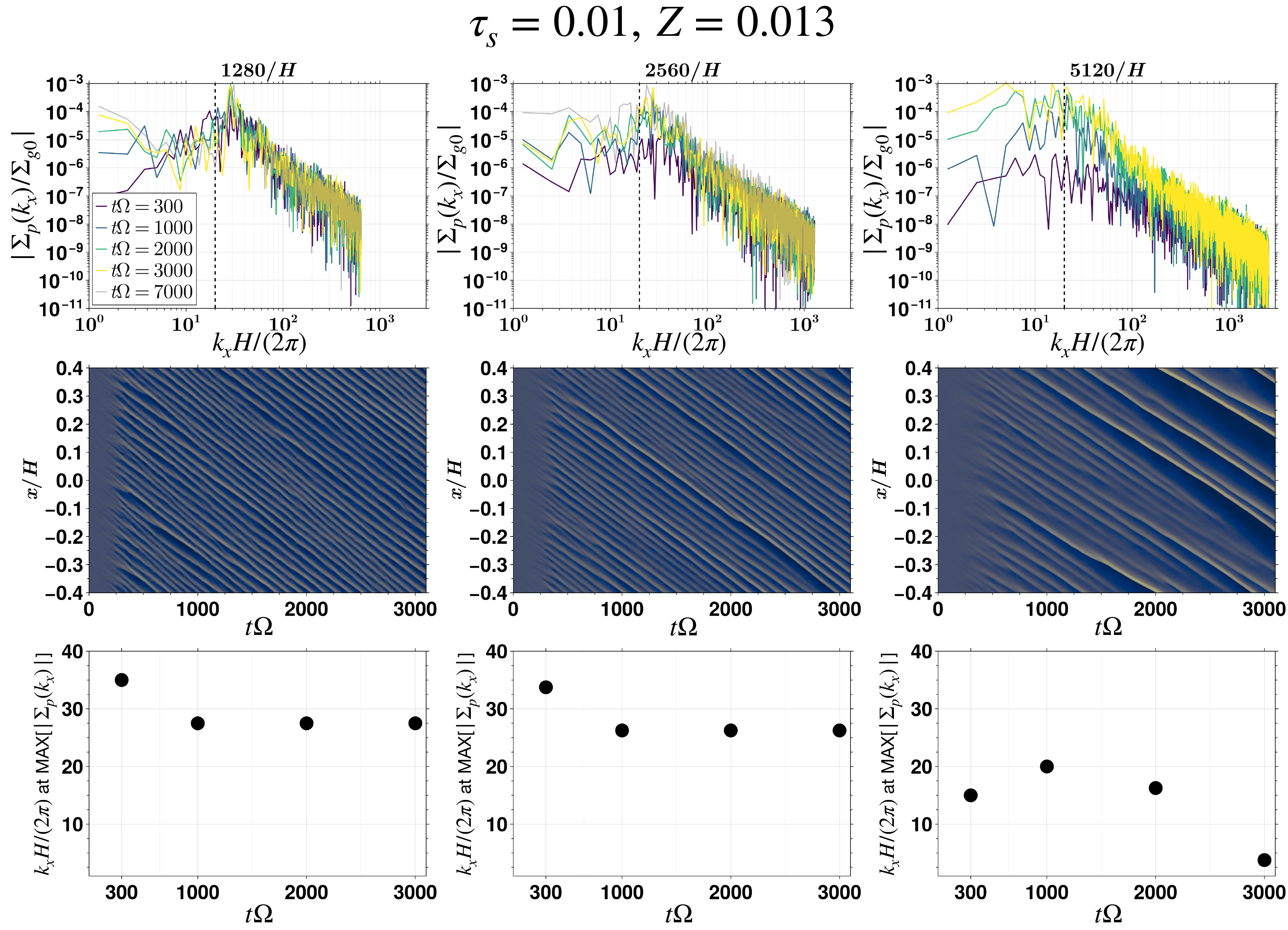}
    \caption{Top: the Fourier power spectrum of particle surface density $(\Sigma_p)$ as a function of radial wavenumber ($k_x$) for $\tau_s = 0.01$, $Z = 0.013$. The spectra are shown at different times by different line colors, with resolutions of $1280/H$, $2560/H$, and $5120/H$ shown in the left, middle, and right panels, respectively. The wavenumber corresponding to the scale of $\eta r$ is marked by the black vertical line in each panel. Middle: spacetime diagrams of $\Sigma_p$ for $t\Omega \in [0,3100]$ at each resolution. Bottom: the $k_x$ associated with the peak of the power spectrum vs time. While the data point for $t\Omega = 7000$ is not included for the $1280/H$ and $2560/H$ cases, the peak occurs at the same wavenumber as at $t\Omega = 3000$. The highest resolution exhibits the most significant density growth across a broader range of wavenumbers over the considered time interval, whereas the lowest resolution does not show density growth.}
    \label{fig:fig-rhopk-Z0013}
\end{figure*}

\begin{figure*}
    \centering
    \includegraphics[width=\textwidth]{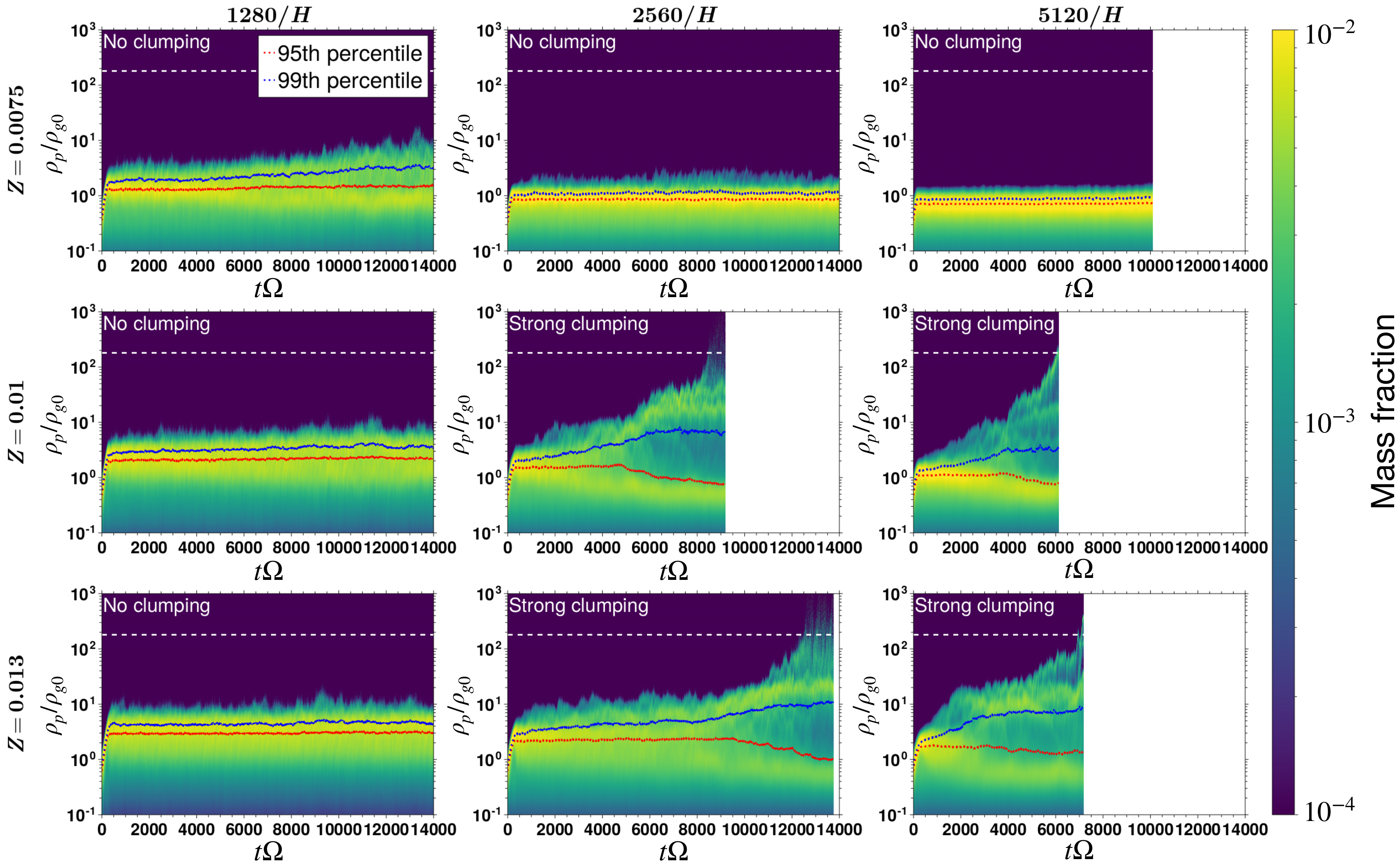}
    \caption{Time evolution of particle histograms, showing the mass fraction as a function of particle density and time for $Z=0.0075$ (top), $Z=0.01$ (middle) and $Z=0.013$ (bottom). All runs shown here have $\tau_s=0.01$. Three different resolutions are shown for each $Z$ value: $1280/H$ (left), $2560/H$ (middle), and $5120/H$ (right). The overlaid red and blue dotted lines in each panel are the 95th and 99th percentiles of the particle density, respectively. The white horizontal line in each panel represents the Hill density (Equation \ref{eq:hill}). Most of the particle mass for the lowest resolution cases is in regions of $\rho_p/\rho_{g0}\sim 1$ (i.e., in weak filaments), whereas that for the strong clumping cases gradually moves towards the region of $\rho_p/\rho_{g0} < 1$. The high density region $(\rho_p/\rho_{g0} > 10)$ is always above the 99th percentile curve. }
    \label{fig:fig-rhophist}
\end{figure*}

\begin{figure*}
    \centering
    \includegraphics[width=\textwidth]{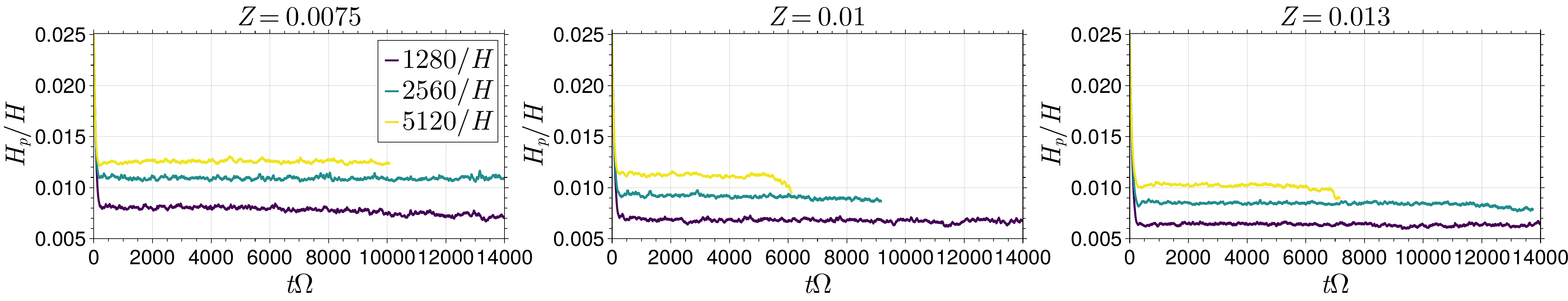}
    \caption{Time evolution of the scale height of particles ($H_p$) for $Z=0.0075$ (left), 0.01 (middle), and 0.013 (right) with $\tau_s=0.01$. In each panel, three different resolutions are color-coded. Regardless of $Z$, the scale height increases with resolution. In the middle and the right panels, the downturn of $H_p$ in the highest resolution case is a result of strong clumps forming.}
    \label{fig:fig-tau001hp}
\end{figure*}

\subsection{Radial Concentration of Particles}\label{sec:result:RadialConcentration}
The SI predominantly clumps particles radially into filamentary structures in both unstratified \citep{johansen_rapid_2007,Baronett24} and stratified (\citealt{YangJohansen14}; \citetalias{Yang2017}; \citealt{Li18}) simulations. We show the radial concentration in the $\tau_s=0.01$ runs in Figure~\ref{fig:fig-rhopt}. The figure shows the particle density integrated over the vertical direction $(\Sigma_p)$ versus time and $x$ (i.e., space-time plots). From top to bottom, the rows correspond to $Z=0.0075, ~0.01$, and 0.013, and from left to right, the columns correspond to grid resolutions of $1280/H$, $2560/H$, and $5120/H$. 

Dense filaments form when $Z \geq 0.01$ and the grid resolution is higher than $1280/H$. Even though filaments also appear in the lowest resolution case (panels on the left column), they are regularly spaced and have relatively low density. In the higher resolution cases, on the other hand, the regular spacing of filaments eventually disrupts. This disruption results in  merging between filaments\footnote{The merging is less noticeable in the $Z=0.01$ cases (i.e., the middle and the right panels in the second row). However, we anticipate that it would occur if the simulations were run for a longer time, given the different drift speeds of the filaments.}, which promotes strong clumping of particles (i.e., maximum particle density exceeding $\rho_H$), leaving fewer but much denser filaments at the end of the simulations. It is important to note that unless $Z$ is sufficiently high for a single filament to reach a maximum particle density comparable to $\rho_H$, merging seems to be necessary for strong clumping based on our simulations.

Interestingly, $Z=0.01$ produces dense filaments earlier than $Z=0.013$ when the resolution is $2560/H$\footnote{Similarly, \citet{yang_diffusion_2018} found fewer but denser filaments forming at $Z=0.02$ compared to $Z=0.04$ in non-ideal magnetohydrodynamic dead zone models (see their Figures 9 and 10).}. This suggests that the disruption and the merging of filaments are stochastic.


The dependence of filament formation on grid resolution is evident, but the nature of this dependence varies between $Z=0.0075$ and the other two values. When $Z=0.0075$ (top three panels), filament formation is suppressed as the resolution increases. We will return to this point in Section \ref{sec:result:HpEps}. In the other two cases, higher resolution leads to the earlier formation of dense filaments. This is particularly noticeable for $Z=0.013$: dense filaments form after $t\Omega \sim 1.2\times 10^{4}$ and $\sim 6000$ for the $2560/H$ and $5120/H$ resolutions, respectively. 

To highlight the importance of grid resolution on SI particle clumping, we compute the power spectrum of $\Sigma_p$ as a function of radial wavenumber $(k_x$). Figure \ref{fig:fig-rhopk-Z0013} shows snapshots of the spectrum for each resolution (left to right) with $Z=0.013$. We consider the time interval from $t\Omega=0$ to $t\Omega=7000$ for the $1280/H$ and $2560/H$ cases and to $t\Omega=3000$ for the $5120/H$ case. The black dashed vertical line in each panel is the wavenumber corresponding to a wavelength of $\eta r$, which is the typical length scale of the SI. We also show spacetime diagrams of $\Sigma_p$ zoomed in between $t\Omega=0$ and 3100 in the middle panels. In the bottom panels, we show the $k_x$ corresponding to the maximum power in the particle surface density  to visualize how the peak shifts over time. We first describe the two lower resolutions cases and then the highest resolution case below.

The $1280/H$ and $2560/H$ cases have the peak of the power spectrum around $k_xH/(2\pi) \sim 30$. The corresponding wavelength is $\sim 0.03H$, which is equivalent to the radial distance between regularly spaced filaments seen in the bottom panels. The power spectra at the intermediate resolution show density growth across $k_xH/(2\pi) \sim 1 - 30$, consistent with the presence of a few, slightly dense filaments, whereas no density growth is seen in the lowest resolution case except at the smallest wavenumbers.

In contrast, the highest grid resolution ($5120/H$) case shows significant density growth both in the power spectra and its spacetime diagram, compared to the other two resolutions. The peak of the power spectrum shifts from $k_xH/(2\pi) \sim 20$ to $\sim 4$ between $t\Omega=1000$ and 3000 (see the bottom right panel in Figure \ref{fig:fig-rhopk-Z0013}). This is consistent with the emergence of several dense filaments that have wider spacing than less dense ones. While the intermediate resolution case shows density growth over a range of $k_x$, the highest resolution case exhibits earlier and broader growth, in line with the faster formation of dense filaments. This may suggest that higher-resolution simulations allow particle density to grow across a broader range of scales, facilitating formation of dense filaments as shown in the bottom right panel of Figure \ref{fig:fig-rhopk-Z0013}. We note that although strong clumping (i.e., maximum particle density exceeding $\rho_H$; see Figure \ref{fig:fig-rhophist}) occurs in our higher-resolution simulations with $Z \geq 0.01$, the formation of dense filaments does not necessarily lead to strong clumping, which remains largely stochastic (see Figures 2 and 3 of \citetalias{Yang2017}).

\subsection{Characteristic Particle Density}\label{sec:result:CharRhop}
The maximum density of particles is a common diagnostic to gauge whether strong clumping occurs. However, we emphasize that the maximum density is inherently stochastic since it may only characterize strong clumping based {|bf } a few, or sometimes a single, grid cell. To provide a more robust measure of particle clumping, we calculate a histogram of the particle density across all grid cells to determine the fraction of particle mass in a given density bin.

Figure \ref{fig:fig-rhophist} shows the time evolution of these particle density histograms (i.e, a ``histogram-time" diagram) for the same simulations shown in Figure \ref{fig:fig-rhopt}. The Hill density is shown as the white horizontal line in each panel, and the overlaid red and blue dotted curves in each panel represent the time evolution of the 95th and 99th percentiles of particle density, respectively. We now address the differences in evolution between strong-clumping and no-clumping cases using this analysis.

In the strong clumping cases, most particle mass resides in the vicinity of $\rho_p/\rho_{g0} \sim 1$\footnote{Note that $\rho_p/\rho_{g0}$ in this subsection differs from the midplane density ratio (Equation \ref{eq:eps}); the former is a local ratio over the whole domain, while the latter concerns only the midplane}. for the first few thousand $\Omega^{-1}$. During this period, multiple weak filaments emerge and evolve without significant density growth, as shown in Figure \ref{fig:fig-rhopt}. The peak in the mass fraction is well traced by both the 95th and 99th percentiles. The density associated with those percentiles corresponds to that of the weak filaments. As filaments grow denser, a divergence in the peak mass fraction occurs, with two separate peaks shifting: one toward lower densities $(\rho_p/\rho_{g0} < 1)$ (95th percentile; red) and the other toward higher density regions $(\rho_p/\rho_{g0} > 1)$ (99th percentile; blue). The peak in the lower density region corresponds to {the regions between filaments (see Figure \ref{fig:fig-rhopt}). The higher density region shows more complex behavior, with the number of peaks varying over time. While the 99th percentile curve increases over time in all cases, it stays around $\rho_p/\rho_{g0} \lesssim 10$, about an order of magnitude smaller than $\rho_H$ (the white horizontal dashed line). Thus, the 99th percentile represents the characteristic density of weaker filaments that are not associated with strong clumping. Additionally, we find that only 0.0001\% to 0.01\% of grid cells have particle density higher than $\rho_H$ in the strong-clumping runs. This indicates that strong clumping occurs in a very tiny fraction of grid cells and that these regions are associated with the densest filaments shown in Figure \ref{fig:fig-rhopt}. 

The no-clumping cases display a much simpler evolution. The highest mass fraction is always around $\rho_p/\rho_{g0} \sim 1$. This is consistent with Figure \ref{fig:fig-rhopt} showing that weak filaments, in which the particle density is $\sim \rho_{g0}$, evolve over time without significant density growth.  Moreover, the 95th and 99th percentiles remain close to each other throughout the simulations due to the lack of strong clumping. 

The maximum density of particles in the $Z=0.0075$ cases becomes lower as the resolution increases, which is consistent with the suppressed formation of filaments at higher resolution (see Figure \ref{fig:fig-rhopt}). Similarly, in the other two $Z$ cases, the lowest resolution attains its highest maximum particle density, around $\sim 10\rho_{g0}$, within the first $\sim 500\Omega^{-1}$ to $1000\Omega^{-1}$ and remains there throughout the duration of the simulations. We will return to this resolution effect on particle clumping in the next subsection where we investigate the particle scale height and the midplane density ratio of particles to the gas.

\begin{figure*}
    \includegraphics[width=\textwidth]{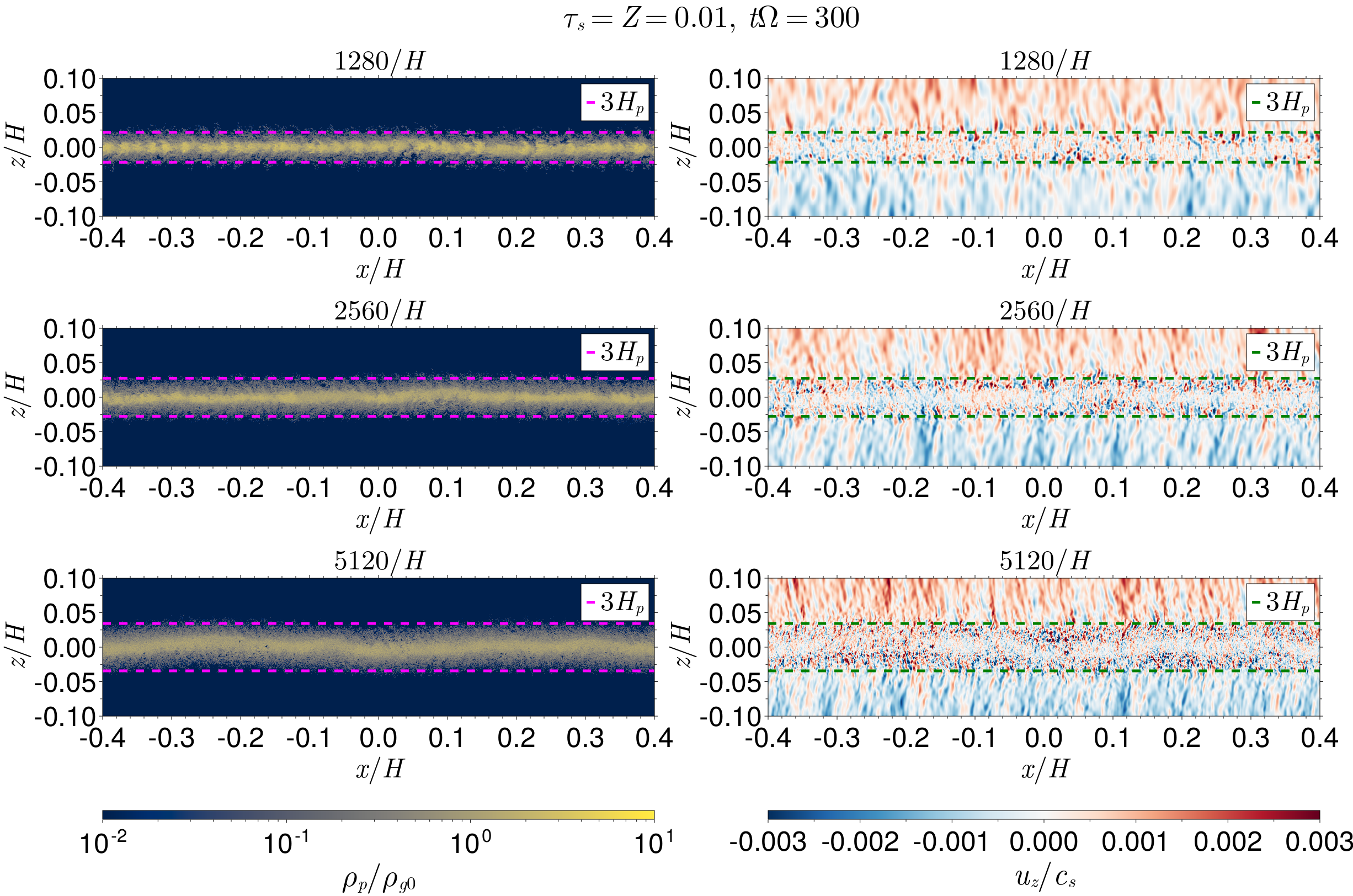}
    \caption{Snapshots of particle density (left) and vertical gas velocity (right) at $t\Omega=300$ in the simulations with $\tau_s=Z=0.01$. Top, middle, and bottom panels correspond to resolutions of $1280/H$, $2560/H$, and $5120/H$, respectively. In each panel, we mark $3H_p$ above and below the midplane with horizontal lines. Higher resolution runs have thicker particle layers, more (and smaller) regions of high velocities near the particle surface layers, and larger velocity magnitudes.  }
    \label{fig:fig-rhop-uz}
\end{figure*}  

\begin{figure}
    \centering
    \includegraphics[width=\columnwidth]{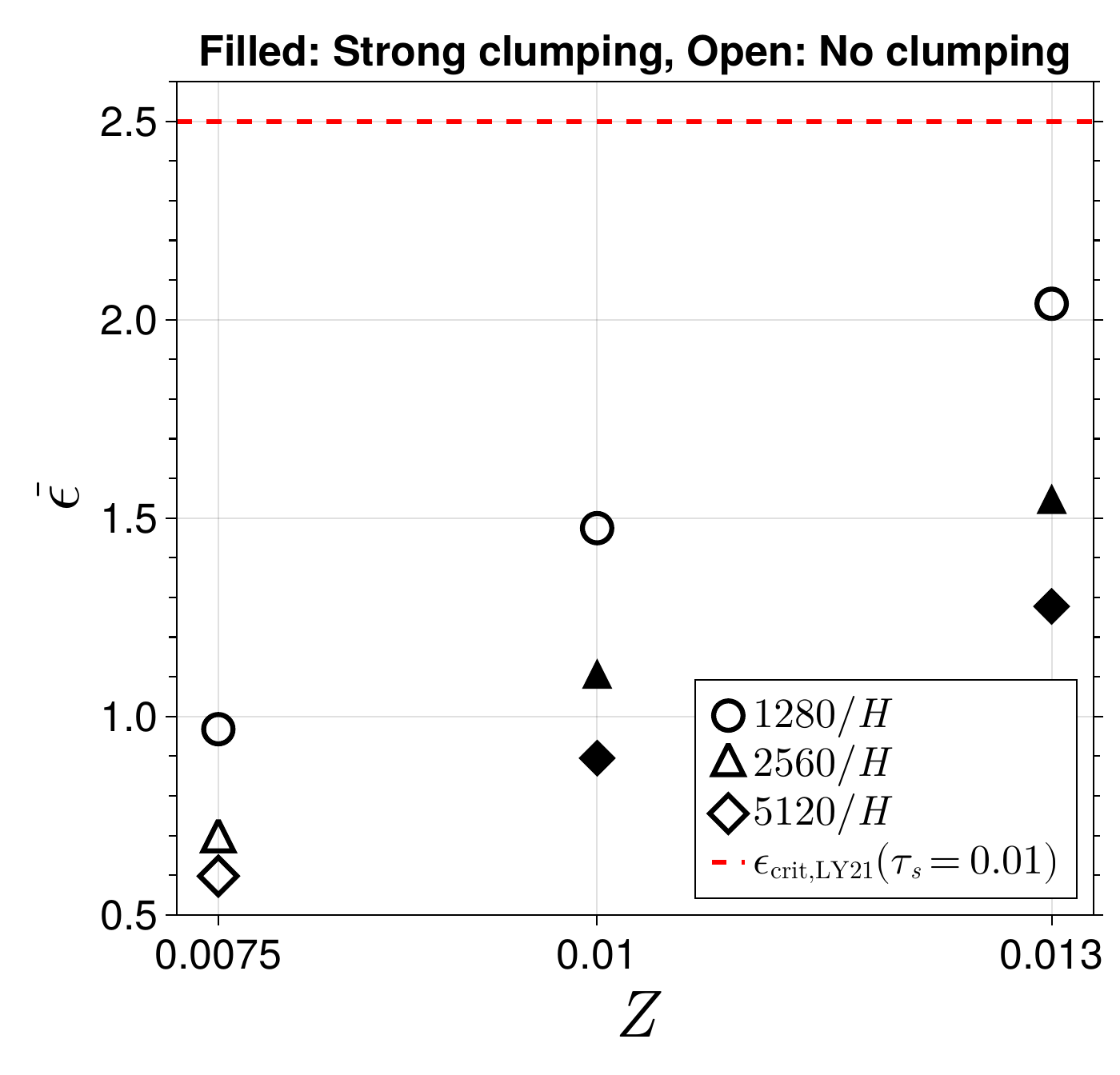}
    \caption{Time-averaged midplane solid-to-gas density ratio ($\overline{\epsilon}$) versus $Z$ for the simulations shown in Figure \ref{fig:fig-tau001hp}. Different resolutions are denoted by different markers: circles $(1280/H)$, triangles $(2560/H)$, and diamonds $(5120/H$). Filled and empty markers correspond to strong and no clumping runs, respectively. The red horizontal line is the critical midplane density ratio, above which strong clumping happens, from \citetalias{LiYoudin21} for $\tau_s = 0.01$ and a resolution of $1280/H$. Even though higher resolution equates to a lower midplane density ratio, strong clumping is triggered for resolutions $\gtrsim 2560/H$ and $Z \gtrsim 0.01$.}
    \label{fig:epstau001}
\end{figure}

\subsection{Scale Height of Particles and Midplane Density Ratio}\label{sec:result:HpEps}
With vertical stratification, particles settle toward the midplane, and if turbulence is present, the particles are stirred by the gas. The competition between sedimentation and gas-driven stirring determines the particle scale height $H_p$ \citep{Dubrulle95,youdin_particle_2007} and the particle-to-gas density ratio at the midplane. As the vertical profile of $\rho_p$ maintains a Gaussian shape in all of our simulations, we calculate the midplane density ratio at a given time during the pre-clumping phase by 
\begin{equation}\label{eq:eps}
    \epsilon  \equiv \left\langle \frac{\rho_p (z=0)}{\rho_g (z=0)} \right\rangle= \frac{Z}{H_p/H}.
\end{equation}
The density ratio is a fundamental parameter, together with $\tau_s$, of the unstratified SI since the linear growth rate for $\tau_s \ll 1$ increases with $\epsilon$, with a sharp increase as $\epsilon$ approaches and exceeds $\sim 1$ \citep{YG05,Youdin_2007}.  However, \citetalias{LiYoudin21} found that the critical midplane density ratio, above which strong clumping occurs, in the \textit{stratified} case varies from $\simeq 0.35$ to $\simeq 2.5$ depending on $\tau_s$. In this subsection, we revisit this critical density ratio, particularly for the $\tau_s = 0.01$ simulations.  However, as a first step, we start with an examination of the particle scale height. 

Figure \ref{fig:fig-tau001hp} shows the time evolution of the particle scale height ($H_p$; see Equation \ref{eq:hp}) for the $\tau_s=0.01$ runs. Moving from left to right, the panels show $Z=0.0075$, 0.01, and 0.013, respectively. In each panel, we color-code different resolutions from $1280/H$ to $5120/H$. The figure shows that $H_p$ increases with resolution for all $Z$ values considered. This effect has also been seen in previous works \citepalias{Yang2017,LiYoudin21}. 

To examine this resolution effect in more detail, we present snapshots of both the particle density $\rho_p$ and the vertical gas velocity $u_z$ (which could  be related to the vertical turbulent diffusion of particles) for the $\tau_s=Z=0.01$ run in Figure \ref{fig:fig-rhop-uz}.  We consider grid resolutions of $1280/H$, $2560/H$, and $5120/H$ in the top, middle, and bottom panels, respectively. The dashed horizontal lines in each panel show $3H_p$ above and below the midplane to denote approximate locations of the surfaces of the particle layers.

Consistent with Figure \ref{fig:fig-tau001hp}, the higher the resolution, the thicker the particle layer. As a result, the particle density at the midplane is higher at lower resolutions (Equation \ref{eq:eps}) in this snapshot. Furthermore, the right panel of Figure \ref{fig:fig-rhop-uz} show that inside the particle layer, the highest velocity magnitudes are reached near the surface of the particle layer. Although this behavior is observed at all resolutions, as resolution increases, the regions in which $|u_z|$ is enhanced (compared to most other regions in the domain) become smaller and more numerous.  There is also a slight increase in the amplitude of these velocity fluctuations with resolution. We confirm that the high velocity fluctuations near the particle surface layers are present for all simulations with $\tau_s = 0.01$.

Given that the vertical gas velocity is largest near the particle surface layers and not at the midplane, it is possible that these gas motions are driven by the vertically shearing streaming instability (VSSI; \citealt{Lin2021}). The VSSI is primarily powered by the vertical shear of the azimuthal velocity of the dust-gas mixture but also needs dust-gas drift. These properties are consistent with our findings considering that the vertical shear is strongest near the surface, where the dust-gas drift is still present (symmetric instability also harnesses the vertical shear but is dynamically a single-fluid phenomenon; see \citealt{Sengupta_Umurhan23}).  That said, we cannot firmly state that it is the VSSI producing this behavior; the VSSI operates at $\sim 10^{-3}H$ scales, whereas our highest grid resolution resolves this scale with only a few grid cells.

Another possible origin for these velocity structures is the Kelvin-Helmholtz instability (KHI).
To examine if the KHI is present in our simulations with $\tau_s=0.01$ (i.e., those presented in Figures \ref{fig:fig-rhopt}-\ref{fig:fig-tau001hp}), we calculated the Richardson number appropriate for axisymmetric KHI ($\textrm{Ri}_r$; see Equation 25 of \citealt{Sengupta_Umurhan23}).  We found that $\textrm{Ri}_r \gg 1$ and $\textrm{Ri}_r$ never gets near the critical value of 0.25, below which the condition for KHI is met\footnote{The critical value of 0.25 is for non-rotating flow (i.e., no rotational forces; \citealt{Chandrasekhar1961,Howard1961,Miles1961}). For differentially rotating disks, previous studies (e.g., \citealt{Johansen06,Lee2010}) suggest a critical value of around unity.}.  This is consistent with results of \citet{Sengupta_Umurhan23}'s axisymmetric simulations with $\tau_s=0.04$ (see their Figure 14). We defer a detailed investigation into the structures and dynamics of this turbulent gas in the future.

We next turn our attention to} the midplane density ratio and its relationship to strong clumping. We calculate the time-averaged value of $\epsilon$ using Equation~\ref{eq:eps} ($\overline{\epsilon}$) as a function of $Z$ (and for three different grid resolutions) and plot the result in Figure \ref{fig:epstau001}. Time averaging is done during the pre-clumping phase. The red horizontal line shows the critical density ratio for $\tau_s=0.01$ found by \citetalias{LiYoudin21} who carried out simulations at the $1280/H$ resolution. Filled and open markers represent strong- and no-clumping cases, respectively. 

At a given $Z$, $\overline{\epsilon}$ decreases with increasing grid resolution. This is a direct consequence of larger $H_p$ at higher resolutions. Although the lowest resolution $(1280/H)$ runs always have the highest density ratio compared to their higher resolution counterparts, none of them produce strong clumping. However, the higher resolutions ($2560/H$ and $5120/H$) with $Z\gtrsim 0.01$ result in strong clumping with $\epsilon$ lower than the critical value proposed by \citetalias{LiYoudin21}. This comparison suggests that $\epsilon_{\rm{crit}}$ depends on grid resolution. While our high-resolution simulations find strong clumping with $\epsilon \sim 1-1.5$, we caution against considering this as the definitive $\epsilon_{\rm{crit}}$ value, as it may change with even higher resolution; achieving convergence may require a grid resolution higher than $5120/H$, which is currently too computationally expensive to explore.

In summary, our results clearly demonstrate that there is an increase in $H_p$ with grid resolution.  While the origin of this effect remains unclear, we speculate that it \text{might be} due to the VSSI. The dependence of $H_p$ on resolution results in $\overline{\epsilon}$ {\it decreasing} with increasing resolution. Despite this, however, higher resolution simulations more readily produce strong clumping even with a reduced $\overline{\epsilon}$. This suggests that increasing resolution has two competing effects: it enhances particle stirring while also promoting particle clumping (as more scales are resolved over which clumping can occur). According to our simulation results, strong clumping occurs when the latter outweighs the former; this happens at $Z \geq 0.01$. In contrast, the $Z = 0.0075$ cases show that increasing resolution suppresses filament formation and results in decreasing maximum particle densities; this indicates that particle stirring is dominant over clumping for this $Z$ value.



\section{Discussion}\label{sec:discussion}

\subsection{Can Linear Theory Predict Strong Clumping?}\label{sec:discussion:linearSI}

Our runs with strong-clumping and $\tau_s = 0.01$ have $\epsilon \sim 1-1.5$  (Figure \ref{fig:epstau001}), which might suggest that the condition of $\epsilon \geq 1$ from the linear, unstratified SI \citep{YG05} applies to the nonlinear, stratified SI as well. However, \citetalias{LiYoudin21} found that strong-clumping and no-clumping runs have nearly identical growth rates in the range of $0.001 \leq \tau_s \leq 1.0$ (see their Figure 8). In addition, they showed that although the linear growth rates of modes captured by their higher resolution simulations (Z1.33t1-2x and Z1.33t1-4x runs) are almost three orders of magnitudes higher than those of the simulations at their fiducial resolution ($1280/H$), the higher resolution runs turned out to be no-clumping cases. These findings highlight an inconsistency between the two distinct regimes: linear, unstratified and nonlinear, stratified.

Nevertheless, since we do find strong clumping in our higher resolution simulations for both $Z=0.01$ and $0.013$, it is worth reevaluating the connection between the linear theory and strong clumping. We revisit this issue by considering different resolutions and $Z$ values at $\tau_s=0.01$. In calculating the linear growth rates, we adopt the same procedure as \citetalias{LiYoudin21}, which is briefly described below.

\begin{figure*}
    \centering
    \includegraphics[width=\textwidth]{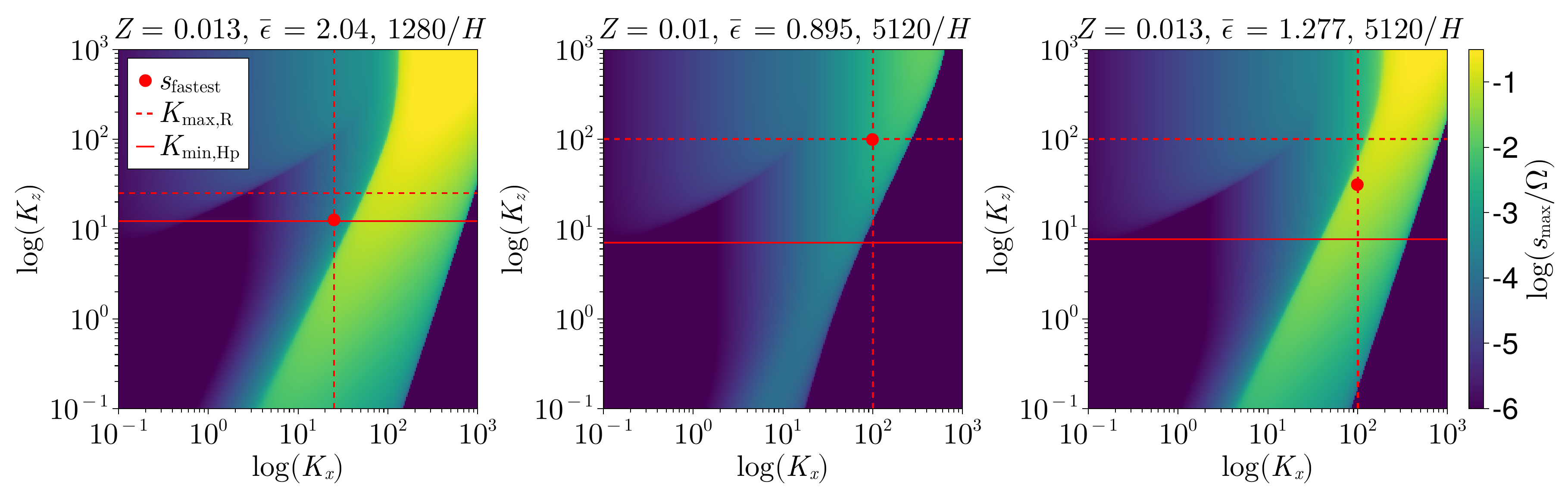}
    \caption{Contours of unstratified SI growth rates as a function of radial ($K_x$) and vertical ($K_z$) dimensionless wavenumbers. The contours shown in the left, middle, and right panels incorporate different simulation parameters into the linear analysis, i.e., different $Z$ and resolutions (all with $\tau_s = 0.01$), as shown by the title of each panel.  When we compute the growth rates, we used $\overline{\epsilon}$ of the corresponding simulation (denoted in the title). For a mode to be resolved in a simulation, the corresponding wavenumber must lie below and to the left of the dashed red lines. The vertical wavenumber of a given mode must lie above the solid red line to fit within the particle layer (see Section \ref{sec:discussion:linearSI} for details of these wavenumber requirements). The red dot in each panel corresponds to the fastest growing mode that satisfy these requirements. The run in the left panel has the highest $\overline{\epsilon}$, but the regions containing the fastest growing modes fall outside of the required wavenumber conditions, whereas the run on the right panel has smaller $\overline{\epsilon}$ but higher $s_{\rm{fastest}}$ is captured within the wavenumber limits due to the higher resolution.}
    \label{fig:growthmap}
\end{figure*}

First, we use dimensionless wavenumbers ($K=k\eta r$) in both radial ($k_x$) and vertical ($k_z$) coordinates. We also assume that for a mode to be properly resolved in simulations, it must be resolved with at least 16 grid cells. This results in a wavenumber requirement for both $K_x$ and $K_z$:
\begin{equation}\label{eq:KmaxR}
    K_{x,z} < K_{\rm{max,R}} = \frac{2\pi \eta r}{R_{\rm{min}}\Delta x} = 2\pi\frac{N_{\eta}}{R_{\rm{min}}},
\end{equation}
where $R_{\rm{min}}=16$, and $N_{\eta} = \eta r /\Delta x$ (i.e., number of cells per $\eta r$). Resolutions of $1280/H$, $2560/H$, and $5120/H$ have 64, 128, and 256 cells per $\eta r$, respectively.

Second, for modes to fit within a particle layer, the vertical wavelength should be no larger than the width of the particle layer. As a result, we demand that half of the maximum wavelength of modes equals $2H_p$. This leads to another requirement for the vertical wavenumber, which is:
\begin{equation}\label{eq:KminHp}
    K_z > K_{\rm{min,H_p}} = \frac{2\pi \eta r}{4H_p} = \frac{\pi}{2}\frac{1}{0.2f_H},
\end{equation}
where $f_H=H_p/(0.2\eta r)$. In calculating $K_{\rm{min,H_p}}$, we use the time-averaged $H_p$ in simulations, as shown in Table \ref{tab:tab2}.

We select 9 runs at $\tau_s=0.01$ (i.e., three different $Z$ and three different resolutions) for the comparison between the linear theory and strong clumping. In each run, we calculate the growth rates of the unstratifed SI by using the time-averaged $\epsilon$ values ($\overline{\epsilon}$; see Table \ref{tab:tab2} and Figure \ref{fig:epstau001}). We present contours of the growth rates of three selected runs in Figure \ref{fig:growthmap}. The title of each panel lists the $Z$ value, $\overline{\epsilon}$, and the resolution of the corresponding run. The vertical and horizontal dashed lines in each panel denote the requirements on the radial and vertical wavenumbers described by Equation (\ref{eq:KmaxR}), respectively. The horizontal solid line in each panel denotes the requirement on the vertical wavenumber (Equation \ref{eq:KminHp}). To satisfy all the requirements, wavenumbers must lie to the left of and below the dashed lines and above the solid line. We find the fastest growth rate in the rectangular region created by these requirements and mark its location as a red circle in each panel. 

\begin{figure}
    \centering
    \includegraphics[width=\columnwidth]{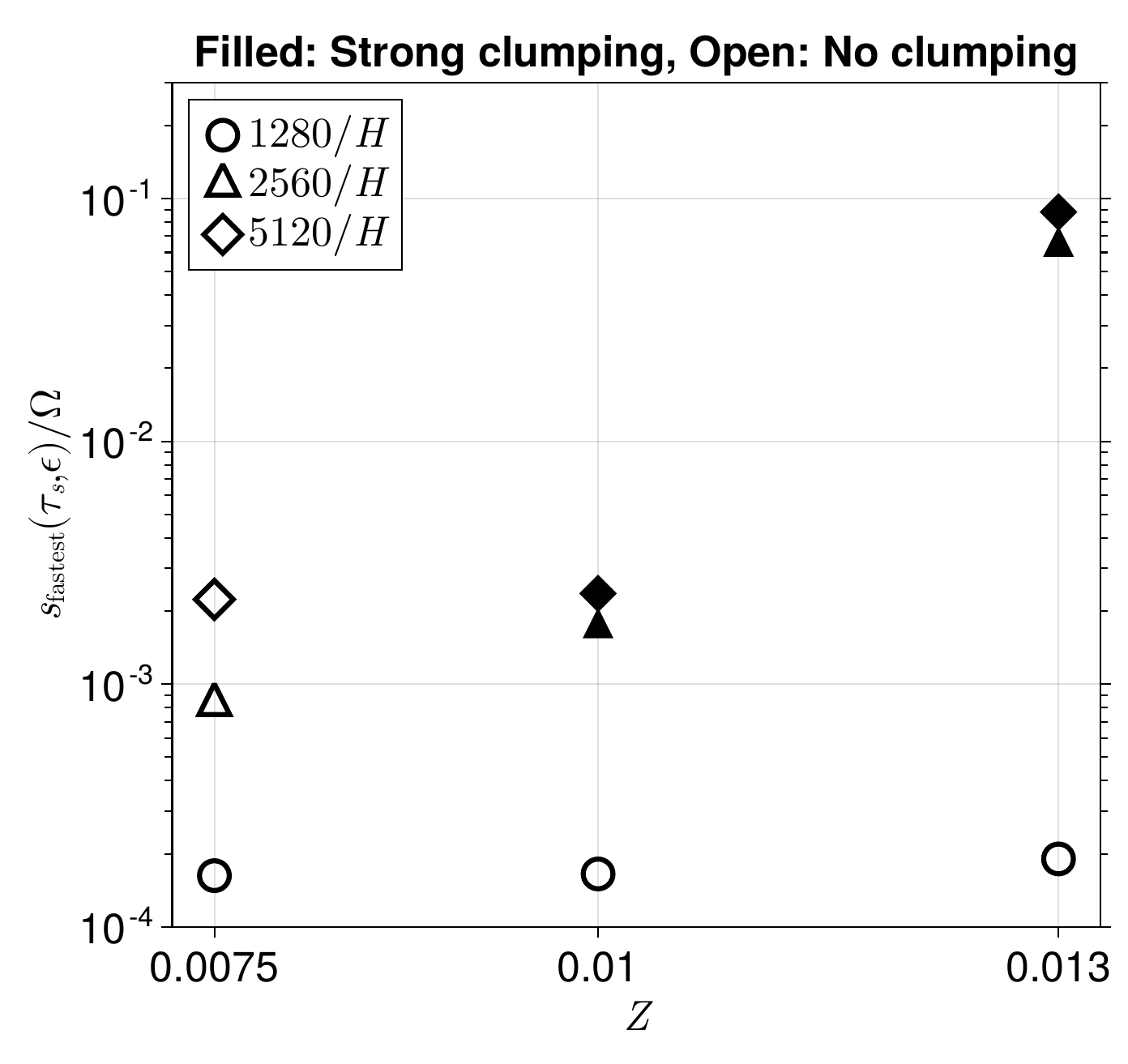}
    \caption{The maximum growth rates that meet the wavenumber requirements illustrated in Figure \ref{fig:growthmap} for runs of three different $Z$ values and three different resolutions.  Different markers are used to denote different resolutions: circles $(1280/H)$, triangles $(2560/H)$, and diamonds $(5120/H$). Filled and empty markers correspond to strong and no clumping runs, respectively. Even though a run with $Z=0.0075$ and a resolution of $5120/H$ has similar fastest growth rate to that of $Z=0.01$ at the same resolution, strong clumping only occurs in the latter, demonstrating that linear growth rates are a poor predictor for nonlinear strong clumping.}
    \label{fig:s_fastest}
\end{figure}

We show the fastest growth rates measured by the aforementioned procedure in Figure \ref{fig:s_fastest}. Circles, triangles, and diamonds denote $1280/H$, $2560/H$, and $5120/H$ resolutions, respectively. We use filled and open markers to denote strong-clumping and no-clumping runs, respectively.

The growth rates increase with increasing resolution for all $Z$ values, even though the lowest resolution cases have the largest $\overline{\epsilon}$ (see Figure \ref{fig:epstau001}). This can be explained by the following: although higher $\overline{\epsilon}$ results in faster growing modes of the linear SI, these faster growing modes occur at higher wavenumbers, which cannot be resolved at the lowest resolution due to the wavenumber restrictions (see also Section 3.5 of \citetalias{LiYoudin21}). To illustrate this, we compare the left and right panels in Figure \ref{fig:growthmap}. The run with $Z=0.013$ and $1280/H$ resolution (left) has higher $\overline{\epsilon}$ than that with $5120/H$ resolution (right). However, the lower resolution cannot probe the high wavenumber regions where \text{the fastest} growing modes reside. On the other hand, the higher resolution case can resolve \text{some of the the fastest} growing modes even though its $\overline{\epsilon}$ is smaller. It is also worth mentioning that the growth rates at the lowest resolution do not vary strongly with $Z$ (circles on Figure ~\ref{fig:s_fastest}), which is consistent with \citetalias{LiYoudin21}.

While the growth rates of the $Z=0.0075$ and $Z=0.01$ runs are similar, this rate significantly increases from $Z=0.01$ and $Z=0.013$ when the resolution is either $2560/H$ or $5120/H$. In fact, $\overline{\epsilon}$ increases from $\sim 1$ to $> 1$ when $Z$ increases from 0.01 to 0.013 at a given resolution, which allows for faster growing modes at $Z=0.013$ (see the middle and the right panels of Figure \ref{fig:growthmap}). In addition, even though the growth rates for the $5120/H$ resolution runs are larger than that of $2560/H$  at a given $Z$, the difference is marginal, especially for $Z=0.01$ and 0.013. This is because the increase in resolution is counterbalanced by the decrease in $\overline{\epsilon}$ (see Figure \ref{fig:epstau001}). Interestingly, the fastest growth rates of the linear SI assuming $\overline{\epsilon}$ values extracted from the strong clumping runs are higher than those of the no-clumping runs for both $Z = 0.01$ and $Z = 0.013$, implying a connection between the linear theory and strong clumping. However, as we now discuss, we find evidence that challenges the connection between the linear theory and strong clumping.

\begin{figure*}
    \centering
    \includegraphics[scale=0.3]{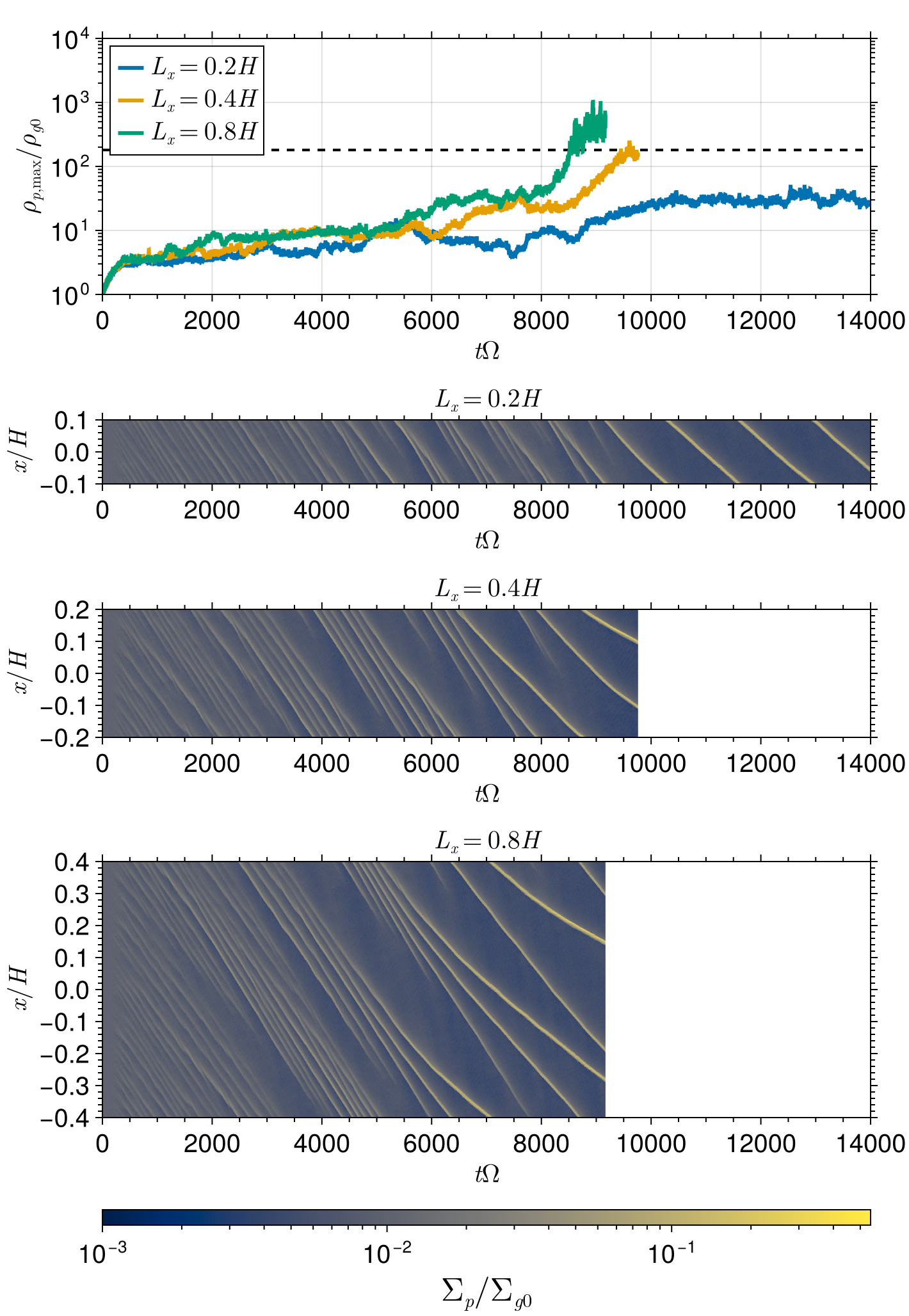}
    \caption{Time evolution of the maximum particle density (top) and space-time plots of $\Sigma_p$ (bottom three panels) for three different $L_x$ values from $0.2H$ to $0.8H$. These runs are initialized with $\tau_s=Z=0.01$, $L_z=0.2H$, a resolution of $2560/H$, and outflow VBCs. The black horizontal line in the top panel is the Hill density (Equation \ref{eq:hill}). Strong clumping occurs for $L_x \geq 0.4H$, with the final number of filaments increasing with $L_x$.}
    \label{fig:Lx-tau001-Z001}
\end{figure*}

First, the linear growth rate does not strongly correlate with whether or not strong clumping occurs. For example, the growth rate for $Z = 0.0075$ at $5120/H$ resolution is nearly the same as that for $Z = 0.01$, yet strong clumping occurs only in the latter. Second, we find that using a larger radial domain (at least $\sim 0.4H$) enhances strong clumping at a fixed grid resolution and $Z$ (Section \ref{sec:discussion:comparison_to_previous_studies}). Since the radial wavelengths of the linear modes (typically $\sim 10^{-2}H$) are much smaller than our domain sizes, changing the domain size should not affect the linear growth at the same resolution. Finally, at a fixed resolution, \citetalias{LiYoudin21} found nearly identical growth rates between no-clumping and strong-clumping cases for $0.001 \leq \tau_s \leq 1.0$ (see their Figure 8). These findings demonstrate that linear, unstratified SI growth rates are not a reliable predictor for strong clumping in stratified simulations. This conclusion does not invalidate either problem but rather demonstrates that the two problems are too distinct to be directly related.

We should note that our analysis has been limited to a comparison with the linear theory of unstratified, {\it inviscid} SI \citep{YG05}. 
While there have been more recent analytical studies \citep{UC20,CL20}, including more complex physics, such as turbulence following an $\alpha$-disk \citep{SS73} formulation (\citealt{YG05} also include a treatment of turbulence in some of their analysis, though not to the same level of detail as these works), vertical stratification \citep{Lin2021}, and even new instabilities that may explain filament formation \citep{GerbigLinLehmann24}, a detailed comparison with each of these studies remains beyond the scope of our current work. For now, however, our primary conclusion in this subsection remains: the linear theory of unstratified, {\it inviscid} SI cannot predict whether or not strong clumping occurs in non-linear stratified simulations.

\subsection{Comparison to Previous Works}\label{sec:discussion:comparison_to_previous_studies}
The clumping boundary (i.e., $Z_{\rm crit})$ has been primarily investigated by \citetalias{carrera_how_2015}, \citetalias{Yang2017}, and \citetalias{LiYoudin21}.
Before proceeding, however, we note that \citetalias{Yang2017} used a different criterion for the clumping boundary compared to \citetalias{LiYoudin21} as well as this work.\footnote{\citetalias{carrera_how_2015} used yet another definition of clumping based on a statistical approach. However, since we will primarily compare with \citetalias{Yang2017} and \citetalias{LiYoudin21}, we opt to not delve into the method of \citetalias{carrera_how_2015} here.} Specifically, \citetalias{Yang2017} examine whether dense particle clumps are visible and persistent in the space-time diagrams of particle density, similar to Figure \ref{fig:fig-rhopt}. We refer to this definition of clumping as ``SI clumping" throughout this section. SI clumping is a more generous threshold than our criterion (i.e., strong clumping) and that of \citetalias{LiYoudin21} since SI clumping can occur even when the maximum particle density remains below $\rho_H=180\rho_{g0}$ (see Figures 2 and 3 of \citetalias{Yang2017}), which is the Hill density for our $Q$ value. However, we note that the Hill density is location-dependent, typically decreasing with increasing radial distance from the star. While this means that the two criteria can be comparable at lower $\rho_H$, we maintain our Hill density requirement through this subsection. 
The first of these studies, \citetalias{carrera_how_2015}, explored the parameter space of $\tau_s$ and $Z$ by gradually increasing $Z$ over the simulation run time for a given $\tau_s$, covering $\tau_s = 0.001-10$.  The two most recent studies, however, performed separate simulations with fixed values of $Z$. Specifically, \citetalias{Yang2017} revisited the clumping boundary, focusing on $\tau_s = 0.001$ and $0.01$ with higher resolutions and longer simulation times. \citetalias{LiYoudin21} conducted simulations across $0.001 \leq \tau_s \leq 1$ and found lower $\Zcrit$ values than both \citetalias{carrera_how_2015} and \citetalias{Yang2017}. All of the previous works used $\Pi=0.05$ as in this work.

\text{In the following two subsubsections}, we focus on comparing our results to the two most recent studies, \text{\citetalias{LiYoudin21} and \citetalias{Yang2017}, respectively} since they perform simulations with fixed $Z$ values as we do here (see section 4.2 of \citetalias{LiYoudin21} for how increasing $Z$ during simulations affects $\Zcrit$). Although \citet{Lim+24} recently examined the clumping boundary using 3D simulations in the presence of forced turbulence, we exclude this study from our comparison as we are not interested in the influence of non-SI-driven turbulence here. 

\subsubsection{Comparison to \citetalias{LiYoudin21}}\label{sec:discussion:comparison_to_previous_studies:LY21}
First, we begin with the comparison to \citetalias{LiYoudin21} by summarizing the main similarities and differences. We use the same $L_x$ as their work but different $L_z$ for the $\tau_s=0.01$ simulations: $L_z=0.2H$ in our simulations, and $L_z=0.4H$ in theirs. Additionally, $n_p=1$ (recall from Section~\ref{sec:methods:parameters} that this is not the effective particle resolution; we use this here instead of $n_p^*$ for simplicity) in our simulations with $L_z=0.2H$, while $n_p=4$ in their simulations. They fixed the grid resolution to $1280/H$ and explored a much larger parameter space, whereas we focus primarily on $\tau_s=0.01$ (though, as discussed above, we do explore some other $\tau_s$ values) to study the effect of resolution on particle clumping for a smaller number of $\tau_s$ values. We consider resolutions up to $5120/H$, which quadruples the fiducial resolution of \citetalias{LiYoudin21}.   

The most interesting difference from \citetalias{LiYoudin21} is that we find strong clumping at $Z=0.01$ and $0.013$, with $\tau_s = 0.01$, whereas they find strong clumping at $Z =0.02$ for that same $\tau_s$, but not for lower $Z$. This results in different $\Zcrit$ curves, with our proposed curve having a much smoother transition between $\tau_s=0.01$ and 0.02 (see Figure \ref{fig:fig-Zcrit}). \citetalias{LiYoudin21} carried out resolution studies at $Z=0.0133$ to examine the sharp transition of their $\Zcrit$ curve by running simulations with $2560/H$ (run Z1.33t1-2x in their paper) and $5120/H$ (run Z1.33t1-4x) but with decreasing box size in the higher resolution simulations: $(L_x,L_z)=(0.2,0.4)H$ and $(L_x,L_z)=(0.2,0.2)H$ for their Z1.33t1-2x and Z1.33t1-4x runs, respectively. They did not find strong clumping at the two higher resolutions; instead, clumping becomes slightly weaker and $H_p$ increases slightly with increasing resolution, the latter of which is consistent with our result (see Figure \ref{fig:fig-tau001hp}). Given that we maintain a larger radial domain size, $L_x=0.8H$, for higher resolution simulations, the difference between our results and \citetalias{LiYoudin21} seems to originate from using different $L_x$ values. To examine this in more detail, we perform two additional simulations at $Z=0.01$ and $2560/H$ resolution, one with $L_x=0.2H$ and the other with $L_x=0.4H$.

\begin{figure*}
    \includegraphics[width=\textwidth]{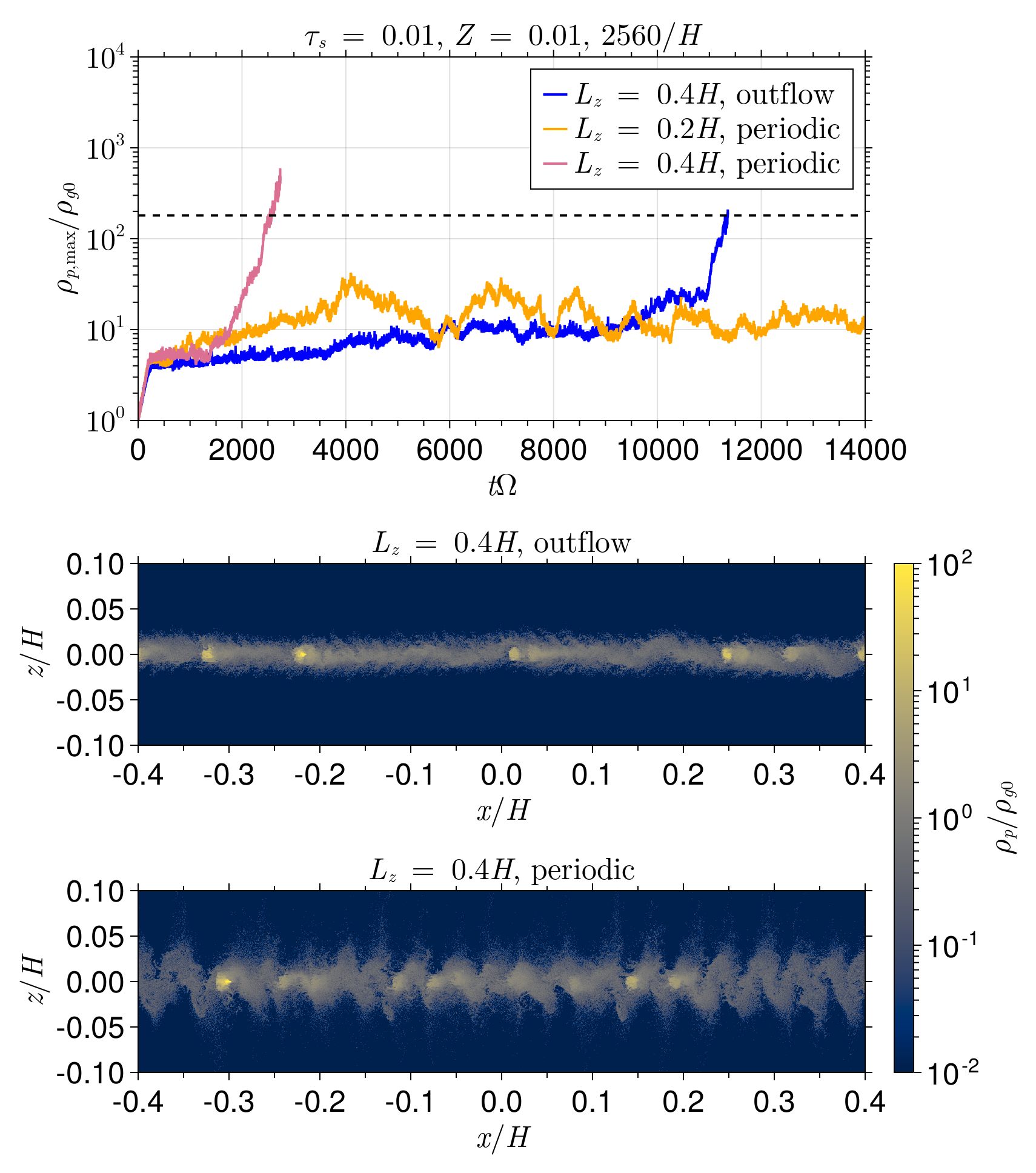}
    \caption{Top: Time evolution of the maximum particle density for three different combinations of $L_z$ and VBCs. Middle: The final snapshot of particle density for $L_z=0.4H$ and outflow VBCs zoomed in to $z/H\in[-0.1,0.1]$. Bottom: Similar to the middle panel but for periodic VBCs. All runs shown here have $\tau_s=Z=0.01$ and $2560/H$ resolution. Strong clumping occurs in the two taller boxes (blue and pink in the top panel), and the run with periodic boundary conditions reaches the Hill density (black dashed horizontal line) much earlier. The snapshots demonstrate that the vertical boundary conditions strongly influence the particle density structure, despite not affecting whether or not strong clumping occurs.}
    \label{fig:Lz-VBC-tau001-Z001}
\end{figure*}

Figure \ref{fig:Lx-tau001-Z001} illustrates the effect of $L_x$ on strong clumping. The upper panel shows maximum particle density as a function of time for three different $L_x$: blue, orange, and green for $L_x=0.2H$, $0.4H$, and $0.8H$, respectively. As can be seen, strong clumping occurs for $L_x \geq 0.4H$, while the maximum density in the smallest $L_x$ case reaches only about $30\rho_{g0}$. 

In the lower three panels, we show space-time diagrams of $\Sigma_p$ for each run with different $L_x$. Multiple filaments form after a few hundreds of $\Omega^{-1}$, and as the simulations progress, some are disrupted and fed into downstream filaments. The final number of filaments is proportional to $L_x$ (see also \citealt{YangJohansen14}): 1, 2, and 4 for $L_x=0.2H$, $0.4H$, and $0.8H$, respectively. In addition, we note that the final maximum density for $L_x=0.2H$ is larger than that for \citetalias{LiYoudin21}'s Z1.33t1-2x run, for which the time-averaged value is $\sim 8\rho_{g0}$. This is because we evolve the model for a much longer time (up to $14,000\Omega^{-1}$), which allows weaker filaments to be disrupted and feed into downstream filaments, ultimately leaving a single, dense filament at the end. In contrast, \citetalias{LiYoudin21} evolved the run up to $3150\Omega^{-1}$ and found several weak clumps at the end of the simulation (see the middle panel of their Figure 11).

This test clearly demonstrates the influence of the mass budget on strong clumping. Even though a single, dense filament forms in the $L_x=0.2H$ run, there is no more material left to feed this filament, preventing the maximum particle density from growing further. On the other hand, when $L_x \geq 0.4H$, there is sufficient mass in the larger domains such that disruptions of weak filaments feed downstream filaments, resulting in strong clumping. Furthermore, in a smaller domain, one can still get strong clumping, assuming $Z$ is sufficiently high (i.e., higher mass budget). 

\subsubsection{Comparison to \citetalias{Yang2017}}\label{sec:discussion:comparison_to_previous_studies:Y17}

We now compare our results to those of \citetalias{Yang2017}. Similar to our study, they considered different resolutions at a given $\tau_s$ and $Z$, focusing on $\tau_s = 0.001$ and 0.01 to assess the clumping boundary. However, they used different domain sizes and VBCs compared to our work. More specifically their simulations used either $L_x = L_z = 0.2H$ or $L_x = L_z = 0.4H$, with higher resolutions employed in the smaller box. Furthermore, they used periodic VBCs, whereas we employ outflow VBCs. Finally, they used $n_p=1$ in every simulation whereas in our cases, $n_p=1$ and 0.5 for $L_z=0.2H$ and $0.4H$, respectively (see Section \ref{sec:method:particle_res}).

\citetalias{Yang2017} found {\it SI clumping} (recall the definition at the beginning of this section) at $Z=0.02$ for $\tau_s=0.01$, with critical resolutions being either $1280/H$ for the larger box or $2560/H$ for the smaller box. However, they did not find any evidence of SI clumping for $\tau_s=Z=0.01$ even though they evolved the simulations for $\sim 2.5 \times 10^{4}\Omega^{-1}$. In contrast, we do see strong clumping after $\sim 6000$--$9000\Omega^{-1}$ depending on resolution (see Figure \ref{fig:fig-rhophist}). This inconsistency may be due to different box sizes and/or the different VBCs used. Since we already addressed the effect of the box size above, we discuss that of VBCs in the following. 

Figure \ref{fig:Lz-VBC-tau001-Z001} compares three different combinations of $L_z$ and VBCs, with fixed $\tau_s=Z=0.01$, and a grid resolution of $2560/H$. The top panel shows the time evolution of the maximum particle density. The maximum particle density of the run with $L_z=0.2H$ and periodic VBCs (orange) levels off around $10 \rho_{g0}$. On the other hand, maximum particle densities of the other two runs with $L_z=0.4H$ (and different VBCs) exceed $\rho_H$ (black horizontal line), triggering strong clumping. 

Strong clumping occurs much earlier with periodic VBCs (pink) than with outflow VBCs (blue), even though outflow VBCs maintain a smaller $H_p$ (\citealt{Li18}, \citetalias{LiYoudin21}; see also the bottom two panels of Figure \ref{fig:Lz-VBC-tau001-Z001}). Considering that the dependence of the scale height on VBCs results from numerical artifacts introduced by the boundary conditions, our finding suggests that strong clumping at this specific parameter choice (i.e., $\tau_s$, $Z$, grid resolution, and domain size) is not significantly impacted by these artifacts. This aligns with \citet{Li18}, who found that the strong clumping is more robust against VBCs in larger boxes by allowing formation of multiple filaments (in radially wider boxes) and reducing artifacts from vertical boundaries (in vertically taller boxes). In addition, it is important to note that this result does not imply that strong clumping always happens faster in periodic VBCs, as the timing of strong clumping---generally caused by disruption and mergers of filaments---may be stochastic.

The middle (outflow VBCs) and the bottom (periodic VBCs) panels in Figure \ref{fig:Lz-VBC-tau001-Z001} show the final snapshots of $\rho_p$ for the two taller box cases. The particle layer is thinner in the outflow case than in the periodic case. Moreover, the morphology of the particle layers are qualitatively different between the two cases. Specifically, the periodic case shows vertically extended structures of low particle density, which is not the case for the outflow VBCs. Since we find a similar structure to the particle layer compared with the simulations of \citetalias{Yang2017} for $\tau_s=0.01$ and periodic VBCs  (see their Figure 1), it is likely that VBCs play a major role in these morphological differences (even though they do not appear to affect whether or not strong clumping occurs).  Finally, the periodic case has fewer clumps than the outflow VBCs. This is consistent with its higher maximum density at the end of the simulation because the two runs have the same $Z$; that is, since the mass budget is the same, the run with fewer clumps may have those clumps reach higher density through mergers.

While the $(L_x, L_z) = (0.8, 0.4)H$ box with periodic VBCs shows strong clumping, it remains unclear whether the lower $\Zcrit$ compared to \citetalias{Yang2017} is primarily due to the larger domain size or the higher resolution. This uncertainty arises because both $L_x$ and the resolution differ from \citetalias{Yang2017}'s highest resolution run $(1280/H)$ with $(L_x, L_z) = (0.4, 0.4)H$ and $\tau_s=Z=0.01$. To address this, we perform another run with the same initial conditions as \citetalias{Yang2017}'s run mentioned above, but at a resolution of $2560/H$, which is twice as high as theirs. 

\begin{figure}
    \centering
    \includegraphics[width=\columnwidth]{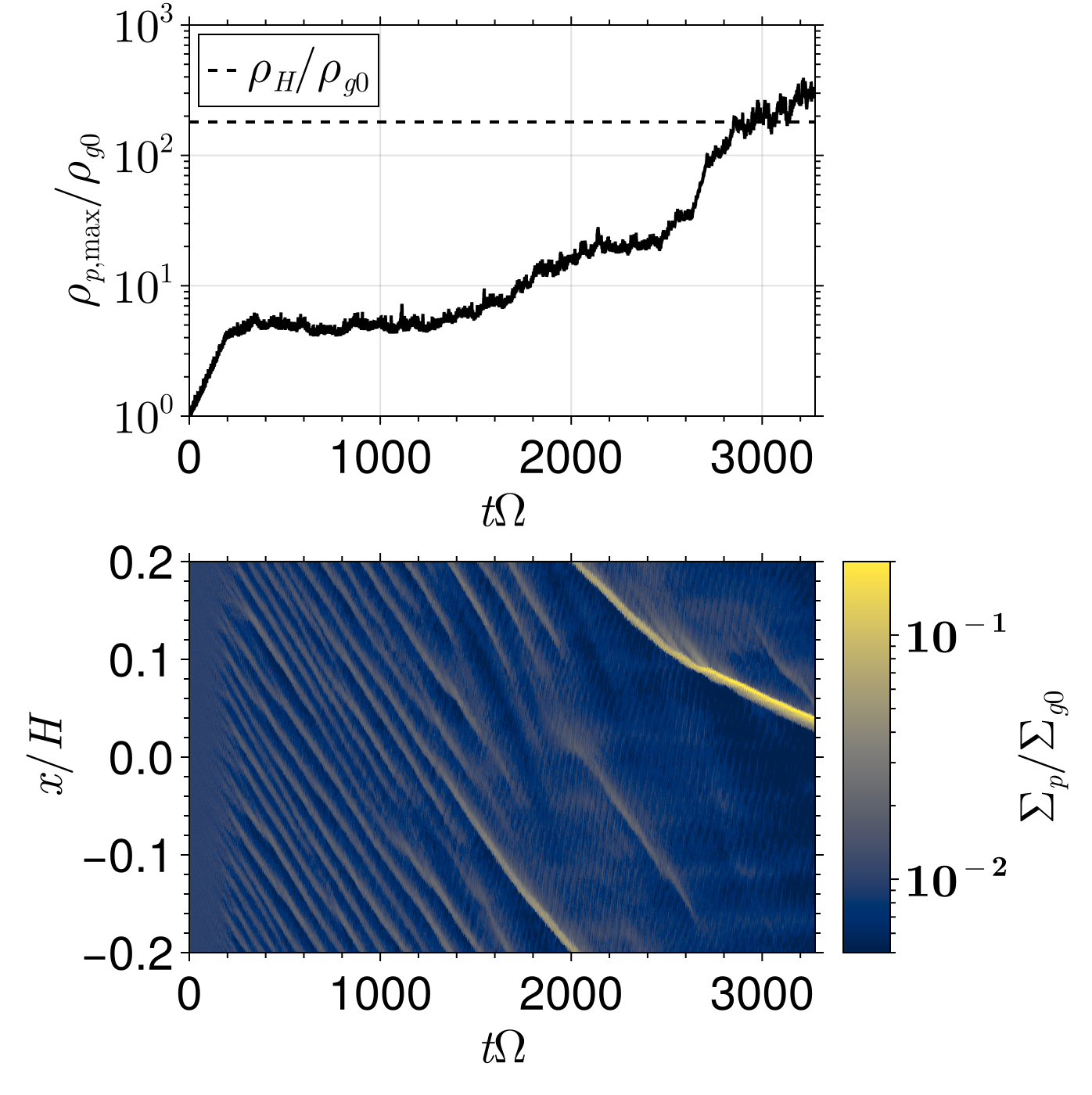}
    \caption{Time evolution of maximum density (top) and the spacetime diagram of $\Sigma_p$ (bottom). The dimensionless stopping time ($\tau_s$) and the surface density ratio $(Z)$ are both 0.01. The run has a domain size of $L_x=L_z=0.4H$ and a resolution of $2560/H$. Periodic boundary conditions are used in the vertical dimension. The maximum density reaches $\rho_H$, thus making this a strong-clumping run, with a single dense filament formed.}
    \label{fig:tau001-Z001-LxLz04H}
\end{figure}

Figure \ref{fig:tau001-Z001-LxLz04H} shows the maximum density of particles over time and the spacetime diagram of $\Sigma_p$ in the top and the bottom panels, respectively, for this new simulation. As the figure illustrates, we observe strong clumping, with a single dense filament forming at $t\Omega \sim 2000$. While this test suggests that increasing resolution is responsible for the transition from no-clumping (as found by \citetalias{Yang2017}) to strong clumping, we note that the maximum density is higher than in \citetalias{Yang2017}’s $Z = 0.02$ run with the same domain size and resolution (see the dotted cyan curve in the bottom-left panel of their Figure 2). This difference likely arises because our run produces a single dense filament, whereas their run exhibits several filaments (see the left panel of their Figure 3). A similar outcome, where lower $Z$ leads to higher maximum density, is also reported in \citet{yang_diffusion_2018}, though they did not delve into the origin of this behavior. Furthermore, \citetalias{Yang2017}'s run with $Z = 0.02$ with the same domain size and resolution as our run presented in Figure \ref{fig:tau001-Z001-LxLz04H} might exhibit strong clumping if the simulation were run for a longer time (they stopped at $t\Omega \sim 6300$), allowing the filaments to merge. This is supported by our $Z = 0.013$ run with $2560/H$ resolution and outflow VBCs (see the bottom-middle panel in Figures \ref{fig:fig-rhopt} and \ref{fig:fig-rhophist}), which shows strong clumping only after $t\Omega \sim 1.2 \times 10^4$.

Additionally, \citetalias{Yang2017} used the {\sc Pencil} code with a very different numerical algorithm for handling stiff drag forces \citep{Yang_Johansen16}. Therefore, we cannot rule out the effect of different codes and algorithms on the strong clumping boundary. While both \Athena ~and {\sc Pencil} codes have been extensively tested, a detailed code comparison is needed to test the numerical robustness of the clumping boundary. However, it is encouraging that we find similar particle layer structure when comparing simulations with identical parameters and VBCs.


Finally, our simulations are 2D axisymmetric. Recent studies have shown that 3D simulations exhibit phenomena that do not occur in 2D simulations, such as apparent outward radial drift of dense filaments \citepalias{Yang2017} and a ``filaments-within-filaments" structure \citepalias{LiYoudin21}. Given these differences, full 3D simulations may alter the clumping boundary presented in this work.

\section{Summary}\label{sec:summary}
In this paper, we examined the conditions for strong particle clumping induced by the streaming instability (SI) in 2D axisymmetric, stratified simulations. We considered three different resolutions, $1280/H$, $2560/H$, and $5120/H$, and large radial domain sizes ($0.8H$ and $0.2H$ in the radial and vertical dimensions, respectively). We define the strong clumping boundary as a critical particle-to-gas surface density ratio $(\Zcrit)$ at a given dimensionless stopping time $(\tau_s)$ above which the maximum density of particles surpasses the Hill density assuming a disk with Toomre $Q$ of 32. Our main findings are as follows:
\begin{enumerate}
    \item According to our simulation results, $\Zcrit$ at $\tau_s=0.01$ is $\sim0.01$, lower than previous works. As such, we provide a new fit to the $\Zcrit(\tau_s)$ curve, which has a smooth transition between $\tau_s~>~0.01$ and $\tau_s \le 0.01$  (Figure \ref{fig:fig-Zcrit} and Equation \ref{eq:Zcrit}). 
    \item We perform an improved statistical analysis of the particle density distribution by calculating the density histogram as a function of time (Figure \ref{fig:fig-rhophist}). We find that the 99th percentile of the density distribution represents the characteristic density of weak particle filaments formed by the SI. 
    \item The particle density at the 99th percentile is about one or two orders of magnitude smaller than the maximum particle density. Furthermore, we found that strong clumping only happens in 0.0001\% to 0.01\% of the total grid cells in our simulations.
    \item We find that the higher the resolution, the higher the scale height of particles. This results in lower particle-to-gas density ratios at the midplane $(\epsilon$) at higher resolutions. 
    \item Nevertheless, strong clumping only happens in the two highest resolution $\tau_s = 0.01$ cases when $Z$ is either 0.01 or 0.013 despite $\epsilon$ being lower than that in the lower resolution runs. 
    \item We confirm previous findings that strong clumping in stratified simulations is not well predicted by the growth rates of the linear, unstratified, inviscid SI. 
\end{enumerate}

Overall, our results demonstrate that the SI can efficiently concentrate dust grains even with the canonical ISM dust-to-gas ratios. Our new $\Zcrit$ curve is a significant refinement on the conditions for planetesimal formation by the SI, highlighting its pivotal role in triggering planetesimal formation in protoplanetary disks. 

\section*{Acknowledgments} 
We thank Min-Kai Lin, Marius Lehmann, Jiahan Shi, Jip Matthijsse, Orkan M. Umurhan, and Debanjan Sengupta for useful discussions. J.L. and J.B.S acknowledge support from NASA under Emerging Worlds grant \# 80NSSC20K0702 and under the Theoretical and Computational Astrophysical Networks (TCAN) grant \# 80NSSC21K0497. J.L. acknowledges support from NASA under the Future Investigators in NASA Earth and Space Science and Technology grant \# 80NSSC22K1322. R.L. acknowledges support from the Heising-Simons Foundation 51 Pegasi b Fellowship.
CCY acknowledges the support from NASA via the Astrophysics Theory Program (grant \#80NSSC24K0133) and the Emerging Worlds program (\#80NSSC23K0653).
The computations were performed using Stampede3 at the Texas Advanced Computing Center using XSEDE/ACCESS grant TG-AST120062 and on Pleiades at the NASA High-End Computing (HEC) Program through the NASA Advanced Supercomputing (NAS) Division at Ames Research Center.

\software{Julia \citep{bezanson2017julia}, Makie.jl \citep{DanischKrumbiegel2021},
          Athena \citep{Stone08,stone_implementation_2010,BaiStone10a,simon_mass_2016}}

\appendix
\counterwithin{figure}{section}
\section{Asymmetric Particle Layer in Outflow Vertical Boundary Conditions}\label{sec:appendixB:IssueOutflow}
\begin{figure*}
    \centering
    \includegraphics[width=\columnwidth]{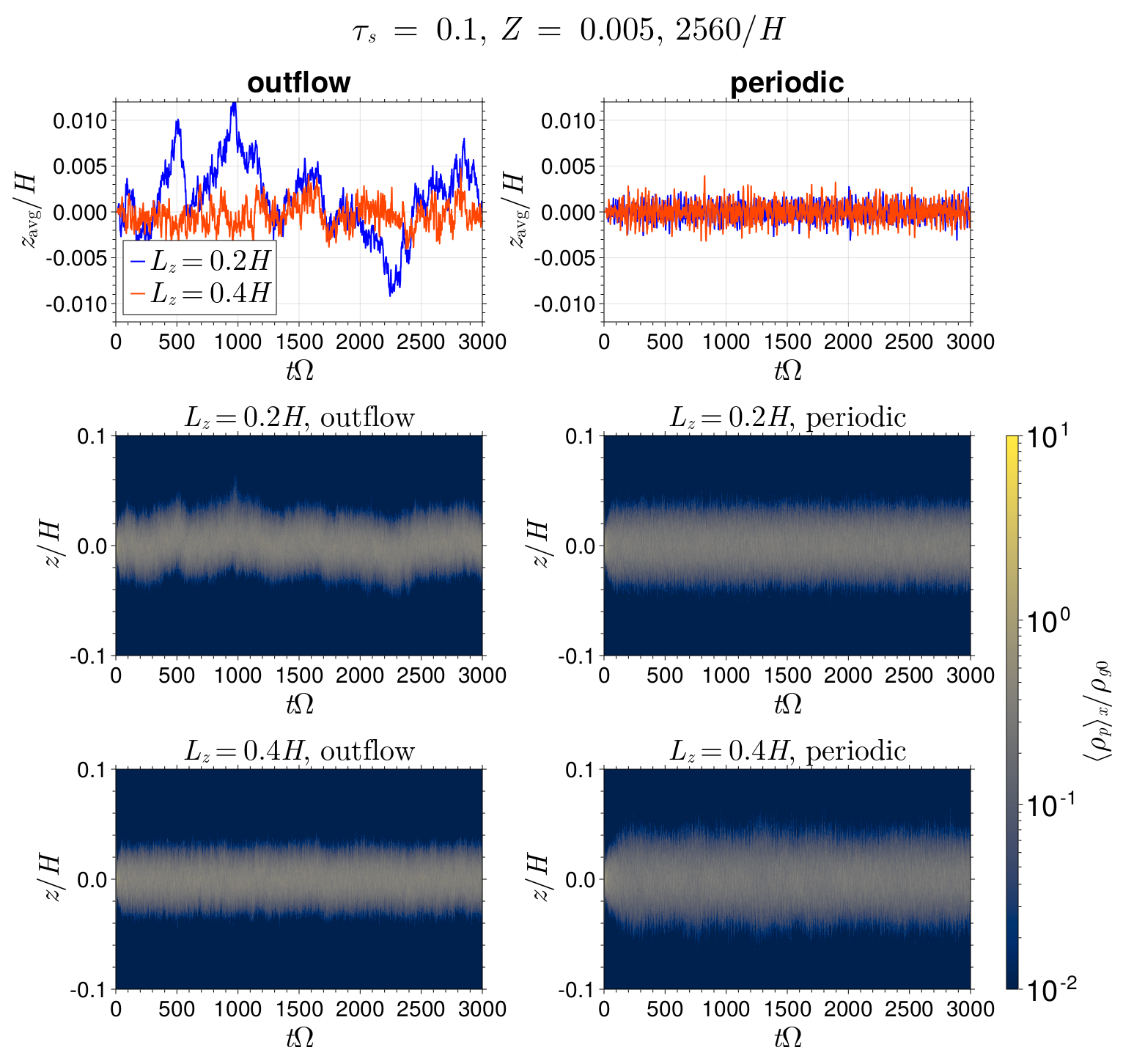}
    \caption{An illustration of the effect of outflow boundary conditions in a short box; two vertical boundary conditions and two different $L_z$ for $\tau_s=0.1, ~Z=0.005$ and $2560/H$ resolution are shown. Left columns correspond to outflow vertical boundary conditions, and the right columns correspond to vertically periodic boundary conditions. Top: mean vertical position of particles versus time for $L_z=0.2H$ in blue and $L_z=0.4H$ in red. Middle: Space-time plots of radially averaged particle density $(\langle \rho_p \rangle_x)$ for $L_z=0.2H$. Bottom: Same as the middle panels but for $L_z=0.4H$. Due to the noticeable asymmetry and minor particle mass loss ($\sim 0.4\%$) in the $L_z = 0.2H$ outflow case, we exclude that simulation from our analysis.}
    \label{fig:figB1-IssueOflow}
\end{figure*}

We select outflow boundary conditions in the vertical dimension as the standard setup because these boundary conditions can mitigate artifacts from the vertical boundaries. This allows more particle sedimentation compared to periodic or reflecting boundary conditions \citep{Li18}. However, in our simulation with $\tau_s=0.1,~Z=0.005$, $L_z=0.2H$, and resolution of $2560/H$, we found that the outflow boundary conditions can lead to significant asymmetry in the particle layer about the midplane. To examine this issue, we compare two different boundary conditions (periodic and outflow) and two different $L_z$ values ($0.2H$ vs. $0.4H$); the results are depicted in Figure \ref{fig:figB1-IssueOflow}, which shows the vertical distributions of particles versus time for outflow (left) and periodic (right) boundary conditions calculated in two separate ways.

First, we calculate the mean vertical position of particles ($z_{\rm{avg}})$ and plot this quantity vs. time in the top panels. Blue and red curves denote $L_z=0.2H$ and $L_z=0.4H$ cases, respectively. When the outflow boundary condition is used, the shorter box shows much more fluctuation in $z_{\rm{avg}}$ than the taller box. On the other hand, the mean position in the runs with periodic boundary conditions is much steadier compared to the outflow VBC runs for both the shorter and the taller boxes. Moreover, its amplitude converges between the two $L_z$ values. 

Second, we compute the radially averaged particle density ($\langle \rho_p \rangle_x(z)$) vs. time as shown in the middle $(L_z=0.2H)$ and bottom panels $(L_z=0.4H)$. The asymmetry of the particle layer is most pronounced in the case of $L_z=0.2H$ with the outflow boundary condition (middle left panel). We also observe that $\sim 0.4\%$ of the total particle mass in the box is lost throughout the simulation. In contrast, the other three cases show nearly symmetric particle layers throughout the evolution. 

Due to this issue, we exclude the simulation with $L_z = 0.2H$ and outflow boundary conditions from our analysis. Additionally, we opt to use $L_z = 0.4H$ for $\tau_s \neq 0.01$ simulations, motivated by \citetalias{LiYoudin21}, who showed that particle scale heights have a local minimum at $\tau_s = 0.01$ and increase toward  $\tau_s < 0.01$ or $0.01 < \tau_s \leq 0.1$ (see their Figure 4). While we use $L_z = 0.2H$ for simulations with $\tau_s = 0.01$, we confirm that no asymmetry issue occurs for this particular $\tau_s$ value. We also find that $\Zcrit$ is robust against different $L_z$ values (see Figure \ref{fig:Lz-VBC-tau001-Z001}).
 
Although the ultimate reason for the significant asymmetry in the shorter box case remains unclear in this study, we speculate that outflow boundary conditions may introduce nonzero net outflows (see Figure \ref{fig:fig-rhop-uz}; see also Figure 1 of \cite{Li18} for the 3D distribution of gas vertical momentum). These outflows may shift the entire particle layer above and below the midplane when $L_z$ is too small. The observed asymmetry of the particle layer in Figure \ref{fig:figB1-IssueOflow} may thus be a result of this effect.

\bibliography{references}{}
\bibliographystyle{aasjournal}

\end{document}